\documentclass[journal]{IEEEtran}
\usepackage{latexsym,epsf,times,amsmath,color,amssymb,indentfirst,subfigure,fancyhdr,colortbl,bm,caption}
\usepackage{graphicx}\epsfxsize=3.in
\usepackage{bbm}

\newfont{\bb}{msbm10 scaled 1100}

\newcommand{\vet}[1]{\mathbf{#1}}
\newcommand{\mat}[1]{\mathbf{#1}}
\newcommand{\artChem}{$\mathcal{AC}$} 
\newcommand{\mol}[1]{\ensuremath{\textrm{#1}}}
\begin{document}

\title{\huge{A Chemistry-Inspired Framework for Achieving Consensus in Wireless Sensor Networks}}
\author{Massimo~Monti, \emph{Student Member, IEEE}, 
Luca~Sanguinetti, \emph{Member, IEEE},
Christian~Tschudin, \emph{Member, IEEE},\thanks{M.~Monti and C.~Tschudin are with the Department of Mathematics and Computer Science, University of Basel, Bernoullistrasse 16, 4056 Basel, Switzerland (e-mail:
\{m.monti,christian.tschudin\}@unibas.ch).} \thanks{L.~Sanguinetti and M.~Luise are with the Dipartimento di Ingegneria dell'Informazione, University of Pisa, Via Caruso 16, 56126 Pisa, Italy (e-mail:
\{luca.sanguinetti, marco.luise\}@iet.unipi.it). L. Sanguinetti is also with the Alcatel-Lucent Chair on Flexible Radio, Sup\'elec, Gif-sur-Yvette, France e-mail:luca.sanguinetti@supelec.fr)} 
and Marco~Luise, \emph{Fellow, IEEE}}
\maketitle

\begin{abstract} 
The aim of this paper is to show how simple interaction mechanisms, inspired by chemical systems, can provide the basic tools to design and analyze a mathematical model for achieving consensus in wireless sensor networks, characterized by balanced directed graphs. The convergence and stability of the model are first proven by using new mathematical tools, which are borrowed directly from chemical theory, and then validated by means of simulation results, for different network topologies and number of sensors. The underlying chemical theory is also used to derive simple interaction rules that may account for practical issues, such as the estimation of the number of neighbors and the robustness against perturbations. Finally, the proposed chemical solution is validated under real-world conditions by means of a four-node hardware implementation where the exchange of information among nodes takes place in a distributed manner (with no need for any admission control and synchronism procedure), simply relying on the transmission of a pulse whose rate is proportional to the state of each sensor.

\end{abstract}
 % \vspace{-0.18in}

%\begin{IEEEkeywords}
%{\bf{To be found.}}
%\end{IEEEkeywords}

%%%%%%%%%%%%%%%%%%%%%%%%%%%%%%%%%%%%%%%%%
\section{Introduction}

The implementation of wireless sensor networks (WSNs) poses several technical challenges. One of primary importance is conjugating the relative unreliability of a single sensor (due to its limited complexity and energy availability) with the high reliability required by certain applications (surveillance, healthcare, factory-automation, in-vehicle sensing and so forth). {For this reason, an intense research activity has been devoted to design algorithms whereby clusters of sensors may reach an agreement on certain quantities of interest in a distributed manner, increasing in this way the system reliability.}
This problem is known in the literature as the \emph{consensus problem} and has received great attention from many different research communities (in computer science, control and information theory, wireless communications and signal processing). {A good survey and treatment of the results obtained in this field can be found in \cite{OlFa07} -- \nocite{AyYi09}\nocite{FrGi11}\cite{DiKa10} and references therein. Although not only limited to these cases, the existing works are basically inspired by different mechanisms (such as biological interactions \cite{Barbarossa2007} -- \cite{Charalambous2010}, formation control \cite{WuGu07}, spreading of gossip in social networks \cite{BoGh06}, synchronization of coupled oscillators \cite{SiSp07}, belief propagation \cite{CoGi06} and so forth), rely on different communication infrastructures (synchronized \cite{AvEl11} or non-synchronized \cite{AyYi09} and \cite{BoGh06}, with or without admission control \cite{SiSp08}\nocite{NoBa11} -- \cite{NaDi11}, clustered-based \cite{GoBo12}, packet-oriented \cite{BoGh06}, and so forth) and might make use of different \emph{a-priori} information (e.g., knowledge about the network topology \cite{DiSa08}). 

{In this work, we focus on WSNs characterized by balanced directed graphs (i.e., graphs in which in-degree and out-degree of each node are the same) and propose to look at the consensus problem through the eyes of a chemist. To this end, we first introduce the key concepts of distributed artificial chemistry, which provide {basic mathematical tools} to (\emph{i}) formalize interactions among distributed nodes in a chemical manner, (\emph{ii}) model the dynamics of the resulting chemical reaction networks in the form of ordinary differential equations (ODEs) and (\emph{iii})~predict the system's equilibrium points. We  then use these tools to solve the problem at hand. 
As we will see, the use of distributed artificial chemistry leads to a consensus model in the same form of that proposed in \cite{OlFa07}. Differently from \cite{OlFa07}, the underlying distributed artificial chemistry lets naturally emerge the interaction mechanisms that are required to drive the dynamical system of each node, in order to operate according to the ODEs. 
Indeed as we will show, distributed artificial chemistry, besides representing a systematic method to design and analyze distributed systems, represents also a powerful tool to define the microscopic interactions and rules that are needed to achieve macroscopic requirements (thanks to the application of basic chemical rules, such as the law of mass action and the conservation principle). 
Furthermore, the use of chemical theory allows making use of new analytical tools, such as steady-state and stability analysis, deficiency zero theorem~\cite{MaHo74} and chemical organization theory~\cite{DiSp07}. In this work, as a first attempt in this direction, the convergence and stability of the derived mathematical model are proven by using the deficiency zero theorem \cite{Fe72}. 

The performance of the chemical interaction mechanisms is validated by means of simulation results under different operating conditions and settings (i.e., different network topologies and number of sensor nodes). Comparisons are made with existing solutions based on gossip protocols. Finally, the underlying chemical theory is exploited to include, in the dynamical system, mechanisms to account for some practical issues, such as the estimation of the number of neighboring nodes as well as perturbations. } 

To validate the proposed solution and to demonstrate that the chemical paradigm is not a mere intellectual exercise, we also present some real measurements obtained from a four-node hardware testbed. In this simple implementation, the information exchange among the spatially distributed sensors takes place by means of the transmission of a pulse whose rate is proportional to the state of each node. The experimental results are in line with analytical and simulation results, and show that the proposed solution performs reasonably well even under real-word conditions. 

The first attempt to use distributed artificial chemistry for achieving consensus can be found in \cite{MeTh} and \cite{MeTs09a}, in the context of packet-oriented communication networks. Differently from \cite{MeTh} and \cite{MeTs09a}, the main strength of this work lies in the comprehensive treatment of how to make use of chemistry-inspired mechanisms for achieving consensus in WSNs.
Specifically, the major contributions of this work are the following:
\begin{itemize}
\item We provide a comprehensive treatment of the chemistry-inspired basic tools to design and analyze distributed interaction mechanisms.
\item We show how such tools can be used for (\emph{i}) constructing a consensus model for WSNs characterized by balanced directed graphs, and (\emph{ii}) proving convergence and stability in the derived system. 
\item Simulation results, obtained with the network simulator OMNeT~4.1~\cite{VaHo08}, are used to validate the analysis under different network topologies and number of sensor nodes; comparisons are made with other traditional gossip-inspired mathematical models.
\item We make use of the underlying chemical theory to account for some practical issues. 
\item We validate the performance of the proposed approach by means of a four-node hardware implementation, which relies on an emergent and simple communication protocol where nodes exchange their data in an asynchronous manner with no need for admission control. To the best of our knowledge, this is the first time that a chemistry-inspired algorithm is built in a hardware testbed and validated under real-word conditions.\end{itemize}

{
The use of chemical theory in designing and studying the consensus problem must be seen as an alternative way to look at the problem itself. This approach represents an unexplored field, which may provide new tools for the analysis and implementation of algorithms but still requires further investigations. This work is meant to provide a first comprehensive treatment on this topic that ranges from theory to (early) implementation, and we hope it may serve as an incentive for the research community for further explorations in this context.
%We conclude this introductory section by remarking the fact that the use of chemical theory in designing and studying the consensus problem must be seen as an alternative way to look at the problem itself. We hope that this work will stimulate the research community to perform further investigations in such an unexplored field that may provide new tools for the analysis and implementation of algorithms. Although limited in scope, this work is basically meant to provide a first solid evidence in this direction. 
%We conclude this introductory section by remarking that the design of consensus algorithms by means of the chemical theory is an unexplored field, which must be seen does not provides necessarily solutions that perform optimally in all respects or better than existing ones. Rather, describing distributed interactions by means of the chemical theory does provide an alternative interpretation of the consensus problem itself, and it provides new tools that may lead to benefits in terms of analysis or implementation. Although limited in scope, our work aims at highlighting some of these aspects and at serving as an incentive to the research community for further investigations in such an unexplored field.
} 

%%-------------------------------------------------------------------
%\subsection{Paper structure}
The remainder of this paper is organized as follows. The system model and consensus problem are briefly introduced in the next section. The key concepts of distributed artificial chemistry are first described and then applied to a simple chemical network in Section~\ref{sec:IntroChem}. The chemistry-inspired consensus model is derived, analyzed, and validated through simulations in Section~\ref{sec:WSN_CNP}. Its extension to account for some practical issues is discussed in Section~\ref{sec:Robustness} and it is again validated with simulations, as well as compared with other gossip-based algorithms. The hardware implementation is illustrated and the related experimental results are reported in Section~\ref{sec:HWimpl}. Additional insights about the chemical approach and some open issues are discussed in Section~\ref{sec:Discussion}.
Finally, some conclusions are drawn in Section \ref{sec:Conclusion}.}

%---------------------------------------------------------------------
\section{Problem statement}\label{sec:probStatement}
%Blocks
{We consider a cluster of $M$ low-mobility sensors\footnote{{We consider WSNs organized in hierarchical levels: the lower level nodes cooperate to achieve local consensus with a reliability greater than the one obtained with a single node; intermediate nodes are responsible for conveying the information gathered by the lower level nodes to the control centers.}} connected by wireless links and composed of the following basic components:}
%We consider WSNs organized in hierarchical levels, where the lower level nodes (cluster) cooperate in a distributed manner to achieve local consensus with a reliability greater than the single node, whereas intermediate nodes are responsible for conveying the local consensus achieved by the lower level nodes to the control centers.} connected by wireless links and composed of the following basic components.}
(\emph{i})~a continuous-time dynamical system whose state evolves
in time according to local measurements and states of
nearby sensors, and (\emph{ii})~a radio transceiver {operating in a half-duplex manner} that is used to transmit
to and receive from nearby sensors. 

The interaction topology of the wireless sensor network is modeled as a {directed graph (digraph)} $\mathcal G = (\mathcal V, \mathcal E)$ in which $\mathcal V = \{\nu_1,\nu_2,\ldots, \nu_{|\mathcal V|}\}$ is the set of all sensors with $|\mathcal V|$ being equal to $M$ while $\mathcal E \subseteq \mathcal V \times \mathcal V$ is the set of edges, with the convention that $(\nu_i, \nu_j ) \in \mathcal E$ if and only if there exists an edge from {$\nu_i$ to $\nu_j$} (i.e., the information flows from {$\nu_i$ to $\nu_j$}). The structure of a digraph can be described by the $|\mathcal V| \times |\mathcal V|$ adjacency matrix $\mathbf {A}$ whose generic entry $[\mathbf {A}]_{i,j}$ is equal to $1$ if $(\nu_i,\nu_j) \in \mathcal E$ and $0$ otherwise.  
As mentioned in the Introduction, we concentrate on balanced digraphs for which the number of edges entering and leaving a node is the same for all nodes, i.e.,
\begin{equation}\label{10.10}
\sum\limits_{j\ne i} [\mathbf {A}]_{i,j} = \sum\limits_{j\ne i}[\mathbf {A}]_{j,i}\quad \forall i \in \mathcal V.
\end{equation}
For notational convenience, we denote by $\mathcal{N}_i$ the neighbor set within the transmitting and receiving range of sensor $\nu_i$, i.e.,
%receiving (listeners)
\begin{equation}
\mathcal{N}_i = \left\{ \nu_j \in \mathcal V \;| \;(\nu_i,\nu_j) \in \mathcal E \right\}.
\end{equation}
Denoting by $z_i$ the {discretized} measurement of sensor $\nu_i$, the goal of this work is to {design, analyze and implement} a dynamical system to distributively calculate {at each node in the network} the average of initial values:
\begin{equation} \label{eq:scope}
z_\textrm{avg} = \frac{1}{ M}\sum\limits_{i =1}^{M} {{z_i}}.
\end{equation}
%% BLOCK DIAG %%%%%%%%%%%%%%%%%%%%%%%%%%%%%%%%
%\begin{figure}
% \vspace{-0.15in}
% \centering
%\centerline{\includegraphics[scale =0.8]{figs/blockDiag}}
%%
%\caption{Composition of a sensor node.}
%\label{fig:blockDiag}
% \vspace{-0.15in}
%\end{figure}

%%%%%%%%%%%%%%%%%%%%%%%%%%%%%%%%%%%%%%%%%
\section{Principles for a chemistry-inspired design of distributed algorithms}
\label{sec:IntroChem}
Consider a vessel in which two molecular species $\mol{S}_1$ and $\mol{S}_2$ (known as reactants) are present and interact {with} each other according to the following rule (reaction): consuming a molecule $\mol{S}_1$ produces a molecule $\mol{S}_2$ and vice-versa. It can easily be proven that the above interaction reaches an equilibrium in which molecules $\mol{S}_1$ and $\mol{S}_2$ are present in the same quantities. This is achieved for any initial concentration of $\mol{S}_1$ and $\mol{S}_2$ and without being explicitly programmed. 

Although simple, such an example (henceforth, called chemical reversible network) shows how an equilibrium can emerge from simple {random interactions whose specific outcomes cannot be easily} predicted. This makes the chemical metaphor suited for the consensus problem in WSNs as they are inherently characterized by high randomness and unpredictability. The challenge to import the chemical paradigm into such networks relies on designing {algorithms, mimicking chemical reactions, so as to enable the achievement of equilibrium points (in accordance with the requirement specification), in a distributed setup}. For this purpose, the key concepts for a chemistry-inspired design and analysis of distributed algorithms are revised in the following section.

%-------------------------------------------------------------------------------
\subsection{Distributed artificial chemistry}\label{sec:ArtChem}
%AC definition
As defined by Dittrich \emph{et al.} in \cite{DiZi01}, an artificial chemistry \artChem{} is a ``\emph{man-made system that is similar to a chemical system}'' in which chemical entities (molecules) interact with each other as specified by abstract models.
%Formal definition of an AC
According to \cite{DiZi01}, an artificial chemistry \artChem{} is univocally defined by the triplet $\mathcal{AC} = (\mathcal{S},\mathcal{R},\mathcal{A})$, where $\mathcal{S}$ is the set of {molecular} species that may appear in a certain chemistry, $\mathcal{R}$ is the set of reaction rules specifying how the molecules interact, and $\mathcal{A}$ is the reaction algorithm describing how and when the reactions are applied. In particular, a reaction rule $r \in\mathcal R$ operates according to a given equation whose general form is as follows:
\begin{equation}\label{eq:reaction}
r: \quad \sum\limits_{s \in \mathcal{S}} a_{r,s} s \mathop \to \limits^{k_r} \sum\limits_{s \in \mathcal{S}} b_{r,s} s
\end{equation}
where $k_r$ {is a constant parameter (known as reaction coefficient) that contributes to regulate the average rate at which reaction $r$ occurs (see also later)}, whereas $a_{r,s}$ is the number of molecules of species $s$ consumed by reaction $r$ (known as stoichiometric reactant coefficient) and $b_{r,s}$ is the number of molecules of species $s$ produced by reaction $r$ (known as the stoichiometric product coefficient). {Basically, the above equation states that reaction $r$ replaces $a_{r,s}$-amount of molecules $s$ to produce $b_{r,s}$-amount of molecules $s$ with an average rate controlled by $k_r$.}

{The dynamics of chemical reactions (when which reaction occurs) are described on average by the well-known law of mass action, which essentially states that the average rate of occurrence of a chemical reaction is proportional to its reactant concentrations \cite{Ab86}}. That is, the more molecules are present, the more likely reactions become. Mathematically, this means that a chemical reaction $r \in\mathcal R$ occurs at a rate $v_r(t)$ proportional to the abundance of involved
reactants:
\begin{equation}\label{eq:LoMA}
v_r(t) = k_r \prod \limits_{s \in \mathcal S} c_s^{a_{r,s}}(t)
\end{equation}
where $c_s(t)$ denotes the molecular concentration of reactant species $s$ at time $t$ {and $k_r$ is the reaction coefficient mentioned before. In artificial chemistry, the law of mass action is respected by properly setting the time instants indicating how and when a generic reaction must be applied.} As mentioned before, such a time setting is handled by the reaction algorithm~$\mathcal{A}$. 
{In all subsequent discussions, we rely on a deterministic version of the algorithm introduced by Gibson and Bruck in \cite{GiBr00}. For completeness, we report algorithm~$\mathcal{A}$ in Appendix A.}

%
%----------------------------------------------------------------------------
%\subsection{Distributed artificial chemistry}
The formal definition of an \artChem{} can be extended to a network with digraph $\mathcal G = (\mathcal V, \mathcal E)$,  {by simply introducing the concept of distributed \artChem{}. Therein, molecules can be} exchanged over the network links by executing reaction rules that generate remote actions. In particular, at each node $\nu_i \in \mathcal V$, a reaction algorithm $\mathcal{A}$ (the same for all nodes) updates a local multiset of molecules according to a set of local reaction rules. That is, each node $\nu_i$ defines a local artificial chemistry as the triplet $\mathcal {AC}_i = (\mathcal M_i, \mathcal R_i, \mathcal A)$ in which $\mathcal M_i = \mathcal{S}_i \cup \mathcal{S}_i^{(j)}$. The set $\mathcal{S}_i$ defines the species of all molecules that can possibly be found in the local multiset, {whereas $\mathcal{S}_i^{(j)} \subseteq \cup_{j\in \mathcal{N}_i}\mathcal{S}_j$} is the set of species that can possibly be found in its neighbors. Each node also defines its own set of reaction rules $\mathcal R_i$ where a reaction $r_i \in \mathcal R_i$ is specified as follows:
\begin{equation}\label{eq:reactDef1}
r_i: \quad \sum\limits_{s \in \mathcal{S}_i} a_{r_i,s} s \mathop \to \limits^{k_{r_i}} \sum\limits_{s \in \mathcal M_i} b_{r_i,s} s
\end{equation}
from which it is observed that all reactants are local species whereas products may also be species located in the neighboring sensors. This is how transmission or exchange of information is modeled in a chemical way: by allowing a reaction to create products in nearby sensors. {Since reactions occur (thanks to the reaction algorithm~$\mathcal{A}$) with an average rate given by the law of mass action in \eqref{eq:LoMA}, it follows that the exchange of information (interactions among nodes) occurs proportionally to the abundance of local reactants.} {Observe also that thanks to the mass conservation principle, if the reaction network forms a closed system, then the total number of molecules is conserved by all reactions and remains constant over time.}

%%%%%%%%%%%%%%%%%%%%%%%%%%%%%%%%%%%%%%%%

\subsection{Dynamical analysis of distributed artificial chemistry}

{The state transition dynamics of a distributed \artChem{} are fully described by the chemical
master equation \cite{Mc67}. Unfortunately, this method becomes too complex in the presence of large reaction networks \cite{JaHu07}. A possible solution is to resort to the method illustrated in \cite{Gi00} in which the mean time evolution of the chemical reaction system is examined. This amounts to looking at the time evolution of the abundance of species, and can be mathematically formalized using the following set of ODEs: 
\begin{equation}\label{eq:ODEs}
{\dot c_{s}} (t)=  { 
	\sum\limits_{r \in {\mathcal{R}}}^{} b_{r,s} v_r(t)
}
 -
{ 
	\sum\limits_{r \in {\mathcal{R}}}^{} a_{r,s}
	 v_r(t) \quad \forall s \in \mathcal S
}
\end{equation}
where $v_r(t)$ is the reaction rate given in \eqref{eq:LoMA}. Denoting by $\vet{v}(t)$ the vector collecting all reaction rates, we may rewrite \eqref{eq:ODEs} in matrix form as follows $ \dot{\vet{c}}(t) = \mat{U} \vet{v}(t)$
where $\mat{U}$ is the stoichiometric matrix whose elements are $[\mat U]_{{s,r}}=b_{r,s}-a_{r,s}$.
{That is, modeling the interactions among nodes according to a distributed \artChem{} allows fully characterizing the system dynamics through a fluid model in the form of ODEs.}
%
%% Distributed reversible game
\subsection{Chemical reversible network}
%%%%%%%
\begin{figure}[t]
 \centering
 \subfigure[Reaction system]{
	\includegraphics[scale = 0.65]{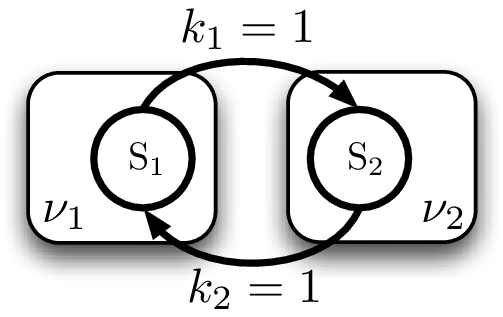}
 \label{fig:revGameReact}
 }
 \subfigure[Simulation measurements]{
	\includegraphics[scale = 0.55]{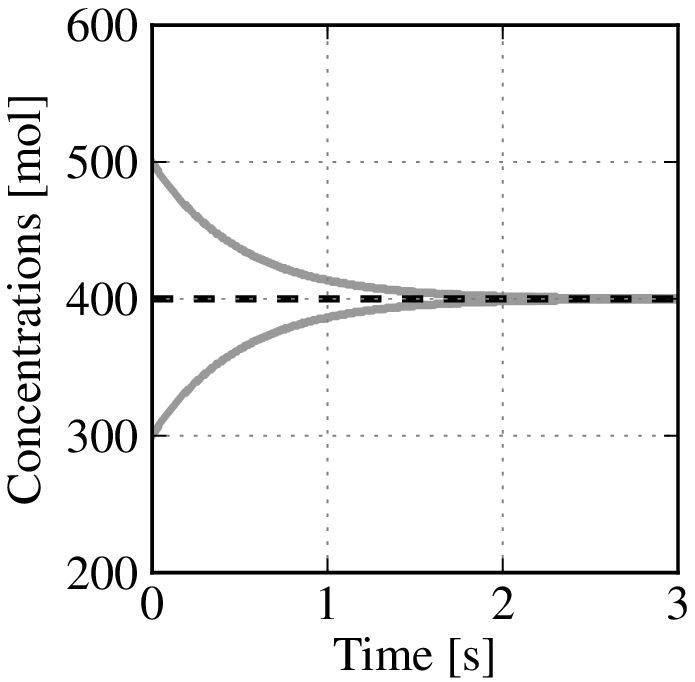}
 \label{fig:revGameRes}
 }
\caption{{Chemical reversible network - The chemical reaction system (a) exhibits an equilibrium point wherein concentrations (gray-continuous line in Fig.~\ref{fig:revGameRes}) converge to the arithmetic mean (black-dashed line in Fig.~\ref{fig:revGameRes}) of their initial concentration values (300 and 500).}}
\label{fig:revGame}
\end{figure}
%%%%%%%
To ease understanding, consider the chemical {network} whose graphical illustration is given in Fig.~\ref{fig:revGameReact}. The network graph is $\mathcal G = (\mathcal V,\mathcal E)$ with $\mathcal V= (\nu_1,\nu_2)$ and $\mathcal E = \{(\nu_1,\nu_2), (\nu_2,\nu_1)\}$. Each node $\nu_i$ defines only a molecular species $\mol{S}_i$ so that $\mathcal{S}_1 = \{\mol{S}_1\}$ and $\mathcal{S}_2 = \{\mol{S}_2\}$. This means that $\mathcal{S}_1^{(2)} = \{\mol{S}_2\}$ and $\mathcal{S}_2^{(1)} = \{\mol{S}_1\}$ whereas $\mathcal M_1 = \mathcal M_2 = \{\mol{S}_1,\mol{S}_2\}$. Additionally, each node defines a single reaction rule that consumes one instance of local $\mol{S}$-molecules to produce one instance of remote $\mol{S}$-molecules. {This leads to the following {``spatially distributed"} reactions: $r_{1}: \mol{S}_1 {\longrightarrow}  \mol{S}_2 $ and $r_{2}: \mol{S}_2	{\longrightarrow}  \mol{S}_1$ }. Collecting all the above facts together yields $\mathcal{AC}_1=\{\{\mol{S}_1,\mol{S}_2\}, r_1, \mathcal A\}$ and $\mathcal{AC}_2=\{\{\mol{S}_1,\mol{S}_2\}, r_2, \mathcal A\}$. From \eqref{eq:ODEs}, we thus have that concentrations change over time according to the following set of ODEs:
\begin{subequations} \label{eq:ChemGameODE}
\begin{align}
	 \dot{c}_{\mol{S}_1}(t) &= c_{\mol{S}_2}(t) - c_{\mol{S}_1}(t)\label{eq:ChemGameODE1}\\
	 \dot{c}_{\mol{S}_2}(t) &= c_{\mol{S}_1}(t) - c_{\mol{S}_2}(t) \label{eq:ChemGameODE2}
 \end{align}
 \end{subequations}
{where we have taken into account that the stoichiometric and reaction coefficients of $r_1$ and $r_2$ are all equal to one.}

When reaching equilibrium at time instant $t=t^\star$, the abundances do not change (i.e., $\dot{c}_{\mol{S}_1}(t^\star) = \dot{c}_{\mol{S}_2}(t^\star) =0$) and the two molecular species are present in the same quantities i.e., $c_{\mol{S}_1}(t^\star) = c_{\mol{S}_2}(t^\star)$. Therefore, by denoting the initial amount of molecules of species $i$ as $c_{\mol{S}_i}(0)$ and {studying \eqref{eq:ChemGameODE} at equilibrium}{,} we obtain 
\begin{equation}\label{EqPointRev}
c_{\mol{S}_1}(t^\star) = c_{\mol{S}_2}(t^\star)=\frac{c_{\mol{S}_1}(0)+c_{\mol{S}_2}(0)}{2}
\end{equation}
{{that proves that the simple} {chemical interaction mechanisms} defined by $\mathcal{AC}_1$ and $\mathcal{AC}_2$ enable to balance the number of molecules between the two nodes. In Fig.~\ref{fig:revGameRes}, we report $c_{\mol{S}_1}(t)$ and $c_{\mol{S}_2}(t)$ as a function of time $t$ when $c_{\mol{S}_1}(0) =500$ and $c_{\mol{S}_2}(0) =300$. The results are obtained in the network simulator OMNeT~4.1 by letting two nodes operate according to the artificial chemistries $\mathcal{AC}_1$ and $\mathcal{AC}_2$. As we can observe, the concentrations of the two species converge to the arithmetic mean of the initial values (black-dashed line).}

{{\bf{Remark}}: Although simple, this example is instrumental to understand what mentioned in the Introduction: modeling network interactions as a distributed artificial chemistry $\mathcal {AC}$ provides (\emph{i}) the microscopic mechanisms (the reactions and their time intervals of execution through reaction algorithm $\mathcal A$) to achieve a specific macroscopic requirement (the average) as well as (\emph{ii}) ODEs that are needed to describe the network dynamics and to eventually compute its equilibrium points.}

% ----------------------------------------------------------------------
\section{A Chemical Consensus Model for \newline Wireless Sensor Networks}\label{sec:WSN_CNP}
As mentioned in the Introduction, the first attempt to make use of distributed artificial chemistry for achieving consensus can be found in~\cite{MeTh}, where the authors propose a chemical networking protocol known as \emph{Disperser}. {The latter is essentially based on the idea of extending the chemical reversible mechanism illustrated in Section III.C to a network with multiple nodes. In particular, each node $\nu_i$ is assumed to contain a molecular species $\mol{S}_{i}$ (i.e., $\mathcal S_i = \{\mol{S}_i\}$) and to randomly interact with one of its neighboring node $\nu_j$ through a spatially distributed reaction $r_{i,j}$. Specifically, reaction $r_{i,j}$ consumes a single $\mol{S}_i$-molecule in the local set of node $\nu_i$ and remotely produces a single $\textrm {S}_j$-molecule in one of the neighboring nodes (i.e., $\mathcal{S}_i^{(j)} \in \cup_{j\in \mathcal{N}_i}\{\mol{S}_i\}$). Mathematically, $r_{i,j}$ is formulated as $r_{i,j}: \; \mol{S}_{i} \to \mol{S}_{j}$.
%\begin{align}\label{r_{i,j}}
%r_{i,j}: \quad \mol{S}_{i} \to \mol{S}_{j}.
%\end{align}}
As basically done for the chemical reversible network, the above interactions are proven to converge towards the average of the initial measurements by simply relying on the mass conservation principle \cite{MeTh}.} 
{From the mathematical expression of remote reactions $r_{i,j}$, it follows that the Disperser} requires each reaction to be associated with a mono-directional link ($r_{i,j}$ consumes a single $\mol{S}_i$-molecule in $\nu_i$ and remotely produces a single $\textrm {S}_j$-molecule in $\nu_j$). This is why the authors in \cite{MeTh} make use of a packet-oriented protocol that basically gives each node the possibility to discern the transmission to and the reception from its neighbors. {Although possible, the use of packet-oriented protocols does not match well the less-demanding communication requirements (in terms of routing and address-managing computation) of WSNs composed by sensors of limited complexity and energy availability.
Following works such as \cite{AyYi09}, \cite{FrGi11}, \cite{NoBa11}, \cite{NaDi11}, we
%{For this reason, following \cite{AyYi09} -- \nocite{AyYi08}\nocite{DoHa11}\nocite{FrGi11}\nocite{AvEl11}\nocite{ZhGo12}\nocite{LiDa07}\cite{DiSa08}\nocite{GoBo12}\nocite{NoBa11}\nocite{NaDi11} in the next we
show next how {to extend the chemical paradigm illustrated above to develop a mathematical model in which the interaction mechanisms take advantage of the broadcast nature of the wireless medium to achieve consensus.}}}
\subsection{Derivation}

{We start setting ${c_{{\mol{S}_i}}} (0) = z_i$ and defining the following ``broadcast" reaction:}%\footnote{{The subscript $B$ stands for broadcast.}}
\begin{equation} \label{eq:remReaction}
	r_{i,B}:  \quad \mol{S}_{i}\stackrel{1}{\longrightarrow}  \sum_{j\, \in \, \mathcal{N}_i} \mol{S}_{j}
\end{equation}
from which it follows that the consumption of a single molecule $\mol{S}_i$ at node $\nu_i$ produces one instance of $\mol{S}$-molecule {at all of its neighbors}. This is exactly how broadcast transmission can be modeled in a chemical way. {According to the law of mass action in \eqref{eq:LoMA}, the above reaction occurs at a rate equal to the concentration of the local state, i.e.,
\begin{equation}
	v_{i,B}(t) = c_{\mol{S}_i}(t).
 \end{equation}
Observe now that the execution of $r_{i,B}$ at each node increases the total number of molecules in the network with no limits, thereby violating the mass conservation principle.} 
%is expected to do. reflects the causes the production of a total of $|\mathcal{N}_i|$ molecules in the system, one for each listening sensor. the reaction reflecting the broadcast transmission between $\nu_i$ and its neighbors $\mathcal{N}_i$ is formalized as given below. If not properly compensated, this would lead to an explosion of the molecular concentrations. 
%correct solution
To overcome this {``diffusion phenomenon"}, we need to further define a reaction that drains the abundance of $\mol{S}$-molecules at each node on the basis of the number of its neighbors. Mathematically, this amounts to locally performing at each node the following {"draining"} reaction:
\begin{equation}\label{eq:drainReaction}
	r_{i,D}: \quad \mol{S}_i  \stackrel{|\mathcal{N}_i| -1}{\longrightarrow} \emptyset. 
 \end{equation}
 {whose rate of occurrence is given by  
\begin{equation}
	v_{i,D}(t) = \left(|\mathcal{N}_i|-1\right)c_{\mol{S}_i}(t)
 \end{equation}
as it follows applying \eqref{eq:LoMA}. Collecting all the above facts together, the artificial chemistry of node $\nu_i$ is { defined as 
\begin{equation}\label{200}
\mathcal{AC}_i=\{\mathcal M_i, \mathcal R_i, \mathcal A\}
\end{equation}
with $\mathcal M_i = \mathcal{S}_i \cup \mathcal{S}_i^{(j)}$, $\mathcal{S}_i = \{\mol{S}_i\}$, $\mathcal{S}_i^{(j)} = \{\mol{S}_j|\,j \in\mathcal{N}_i\}$ and $\mathcal R_i = \{r_{i,B},r_{i,D}\}$}.

}
%------------------------------------------------
%\subsection{Convergence analysis}
Then, from \eqref{eq:ODEs} using \eqref{eq:remReaction} and \eqref{eq:drainReaction} we have that the ODEs describing the evolution of $c_{{\mol{S}_i}}$ take the form: 
\begin{equation}\label{eq:ODE_0}
{\dot c_{{\mol{S}_i}}} (t) =  {\sum\limits_{j \in {\mathcal{N}_i}}^{} {{c_{{\mol{S}_j}}}}(t)} -  {{|\mathcal{N}_i|}{c_{\mol{S}_i}}(t)},\;\;c_{\mol{S}_i}(0)=z_i.
\end{equation}
Recalling that $[\mathbf {A}]_{i,j} = 1$ for any $j \in {\mathcal{N}_i}$ and observing that 
\begin{equation}
{|\mathcal{N}_i|} = \sum\limits_{j \in {\mathcal{N}_i}} [\mathbf {A}]_{i,j}
\end{equation}
we may rewrite \eqref{eq:ODE_0} as
\begin{align}\label{eq:ODE_0_1_a}
{\dot c_{{\mol{S}_i}}} (t) =  \sum\limits_{j \in {\mathcal{N}_i}} {[\mathbf {A}]_{i,j}\left({ c_{\mol{S}_j}(t) - }{c_{\mol{S}_i}}(t)\right)}
\end{align}
or, equivalently, in matrix form
\begin{align}\label{eq:ODE_0_1}
{\dot {\mathbf{c}}_{{\mol{S}}}} (t) =  -\mathbf{L}{\mathbf{c}_{{\mol{S}}}} (t)
\end{align}
where $\mathbf{L}$ is {Laplacian matrix} of $\mathcal{G}$ and $\mathbf{c}_{{\mol{S}}}$ is the vector collecting all species concentrations (nodes' state).
The latter is exactly in the same form of the mathematical model proposed in \cite{OlMu04} and is known to converge towards the average of the initial measurements as formulated in \eqref{eq:scope} when $\mathcal G$ is a strongly connected and {balanced digraph}. %\footnote{{Species $\mol{S}_i$ is initialized to the value of the initial measurement at node $\nu_i$, i.e. $c_{\mol{S}_i}(0)=z_i(0)$.}}
%Differently from \cite{OlMu04}, in which the above fluid model tries to approximate the algorithm dynamics, the set of ODEs in \eqref{eq:ODE_0} is automatically extracted from the reaction network that defines the sensor interactions. 
{Therefore, mimicking sensor interactions in WSNs through distributed artificial chemistries has given us the tools to derive a mathematical model whose convergence to the average is guaranteed in the investigated scenario. {In contrast to} \cite{OlMu04}, the underlying chemical theory allows making use of analytical tools never used in this context before. Indeed, in Appendix B we show how the deficiency zero theorem can be used to prove the convergence and stability of \eqref{eq:ODE_0}. To our knowledge, this is the first time that such a tool is used for proving the convergence of consensus algorithms.}

\subsection{Simulation results}

{{In this subsection}, the performance of the consensus model in \eqref{eq:ODE_0_1} is validated under different operating conditions (different network topologies and number $M$ of sensors\footnote{Due to space limitations, we cannot provide a complete numerical analysis of all the investigated settings. However, all available results will be provided upon request.}) by means of simulation results obtained with the network simulator OMNeT~4.1.} 
{Observe that, thanks to the underlying chemical framework, the simulation of \eqref{eq:ODE_0_1} requires only to let the dynamical system of each node $\nu_i$ operate according to the artificial chemistry {defined in \eqref{200}}. It is worth observing that no synchronous models and admission control mechanisms are required by $\mathcal{AC}_i$. Sensor interactions take place through reactions that are applied according to the time instants of the reaction algorithm $\mathcal A$, driven by the law of mass action. This makes the implementation of the dynamical system a simple task.}

% Fig.  Nets 25 nodi %%%%%%%%%%%%%%%%%%%%%%%%%%%%%%%%%%%%
\begin{figure}[t]
 \centering
 \subfigure[Ring network]{
	\includegraphics[scale = 0.585]{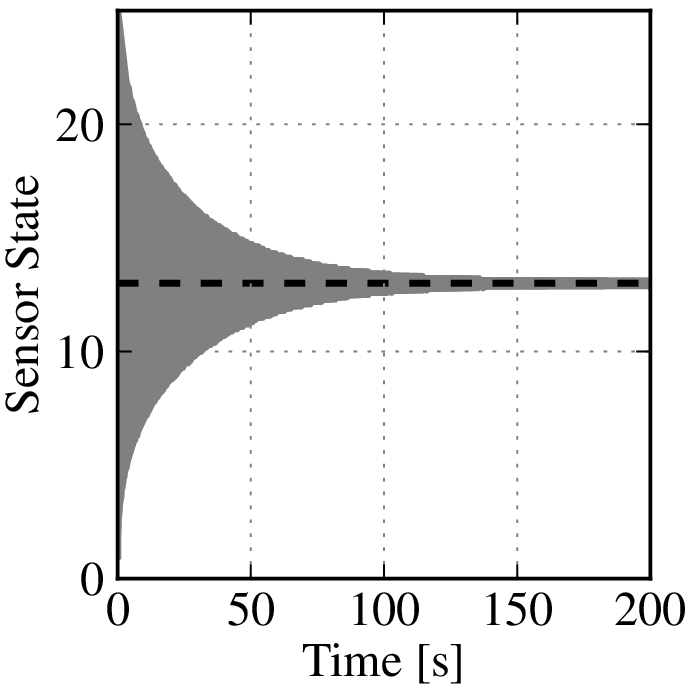}
 \label{fig:ring25}
 }
  \subfigure[{Complete network}]{
	\includegraphics[scale = 0.585]{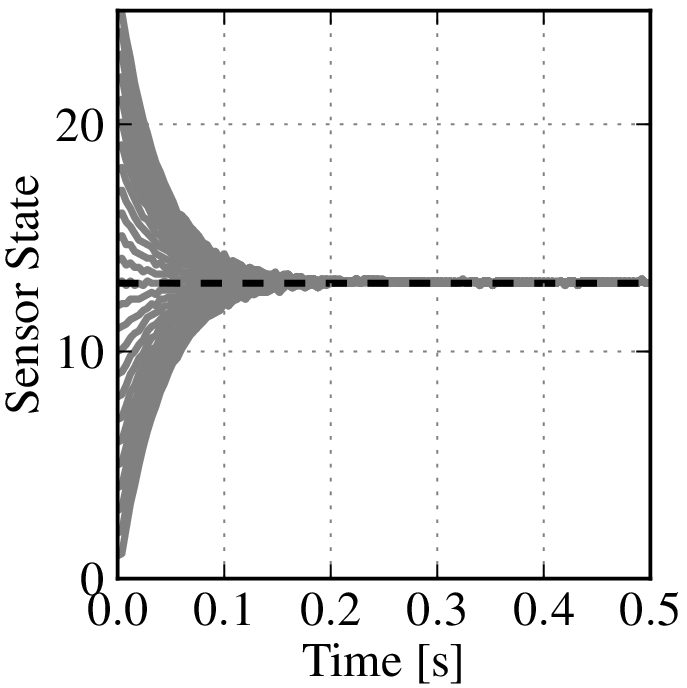}
 \label{fig:full25}
 }
 \subfigure[Regular lattice $k=3$]{
	\includegraphics[scale = 0.585]{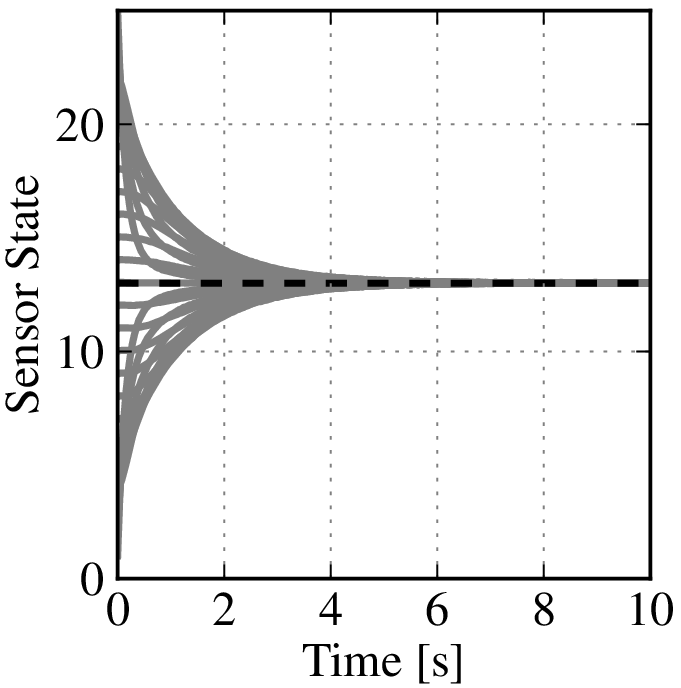}
 \label{fig:rl25}
 }
  \subfigure[{Small-world network}]{
	\includegraphics[scale = 0.585]{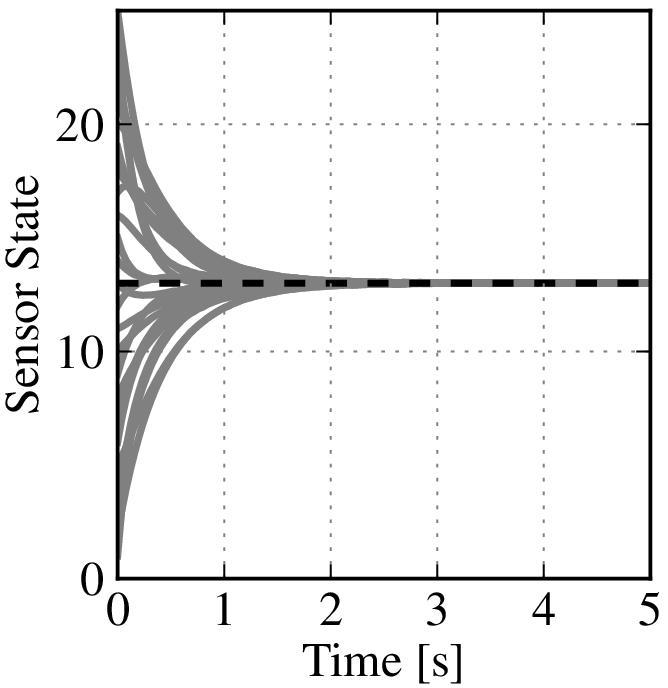}
 \label{fig:sw25}
 }
\caption{{Sensors' state evolution obtained when $M=25$ sensors are connected through (a) a ring network, (b) a complete network, (c) a regular lattice network topology with interconnections to $k=3$ nearest neighbors, and (d) a small-world network topology with $3M$ links (see \cite{OlFa07} and references therein for more details on such networks). The initial state is set to $z_i = i$ for $i=1,2,\ldots,M$.}}
\label{fig:forComparison_2}
\end{figure}

% SmallWorld100 and OlfatiSaber Fig.4.c %%%%%%%%%%%%%%%%%%%%%%%%%%%%%%%%%%%%
\begin{figure}[t]
 \centering
 \subfigure[Regular lattice $k=3$]{
	\includegraphics[scale = 0.585]{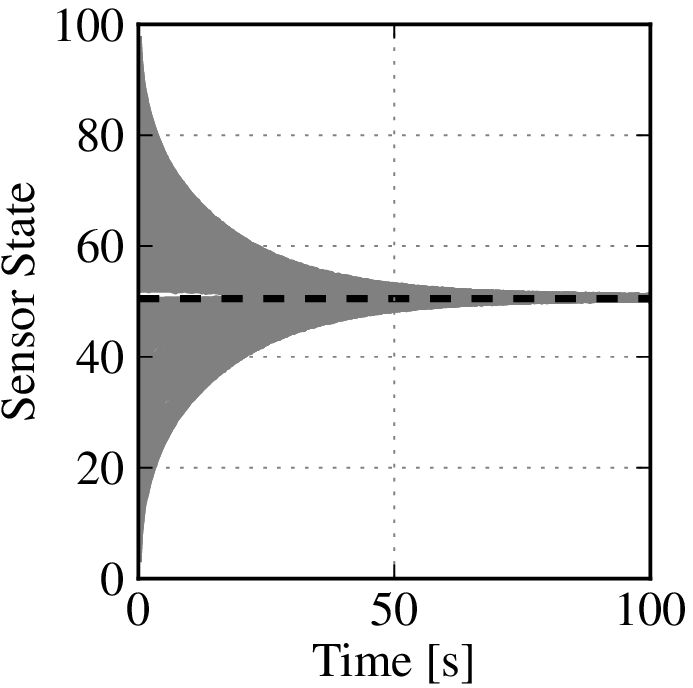}
 \label{fig:comparison}
 }
  \subfigure[{Small-world network}]{
	\includegraphics[scale = 0.585]{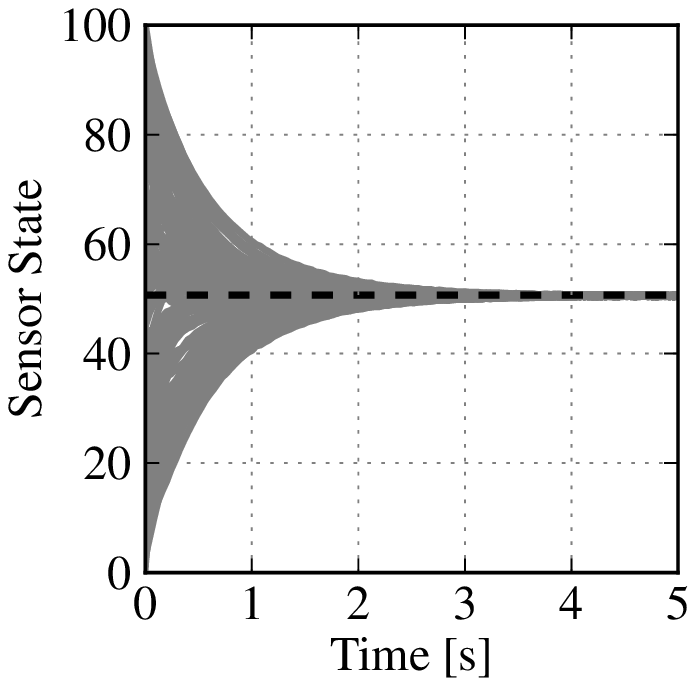}
 \label{fig:100smallNet}
 }
\caption{{Sensors' state evolution obtained when $M=100$ sensors are connected through (a) a regular lattice network topology with interconnections to $k=3$ nearest neighbors and (b) a small-world network topology with $3M$ links. The initial state is set to $z_i = i$ for $i=1,2,\ldots,M$.}}
\label{fig:forComparison}
\end{figure}

%% Ring1000 and SmallWorld1000%%%%%%%%%%%%%%%%%%%%%%%%%%%%%%%%%%%%
%\begin{figure}[t]
% \centering
% \subfigure[Ring network]{
%	\includegraphics[scale = 0.585]{figs/py/bigFigure/newfig/StepRing1000_zoom}
% \label{fig:ring1000}
% }
% \subfigure[Small-world network]{
%	\includegraphics[scale = 0.585]{figs/py/bigFigure/newfig/1000/small1000_3g3p_step_zoom}
% \label{fig:smallworld1000}
% }
%\caption{{Sensor local state evolution with (a) a $M=1000$ ring network topology (b) a $M=1000$ small world network with $3000$ links. The quantities are to $z_1 = 30$ and $z_i= 15$ for $i=2,3,\ldots,M$.}}
%\label{fig:forScalability}
%\end{figure}

%% Convergence table %%%%%%%%%%%%%%%%%%%%%%%%%%%%%%%%%%%%%%%%%
\begin{table*}
 \caption{{Convergence times for achieving a normalized mean squared error less than $0.01$ with different network topologies.}}
 \centering\small
 \begin{tabular}{ | l | r r r r r  |} 
 \hline 
							&$M = 25$ 	&  $M= 50$ 	& $M= 100$ 	& $M= 250$	 & $M= 1000$	
 \\ \hline 
 Ring	 network						&80{\,s}	 	& 230{\,s} 	& 450{\,s} 	& 900{\,s} 		& 1250{\,s}
  \\
 %Regular lattice with $k=3$				&6	 & 25 	& 80 	& nd 	& nd
 Regular lattice with $k=3$				&4{\,s}	 & 12{\,s} 	& 35{\,s} 	& 90{\,s}	& 180{\,s} 	  	
 \\
  Small-world network with $3M$ links	 &1{\,s}	 	& 1{\,s}		 & 1{\,s} 	& 1{\,s}	 	& 1{\,s} 	
 \\ 
Complete network					&0.26{\,s}	 & 0.09{\,s} 	& 0.04{\,s} 	& 0.012{\,s} 	& 0.005{\,s} 	
 \\ 
 \hline 
 \end{tabular}
 \label{tabTime}
\end{table*}

%%%%%%%%%%%%%%%%%%%%%%%%%%%%%%%%%%%%%%%
% RUN%%%%%%%%%%%%%%%%%%%%%%%%%%%%%%%%%%%%
\begin{figure}[t]
 \centering
  \subfigure[Ring network]{
	\includegraphics[scale = 0.585]{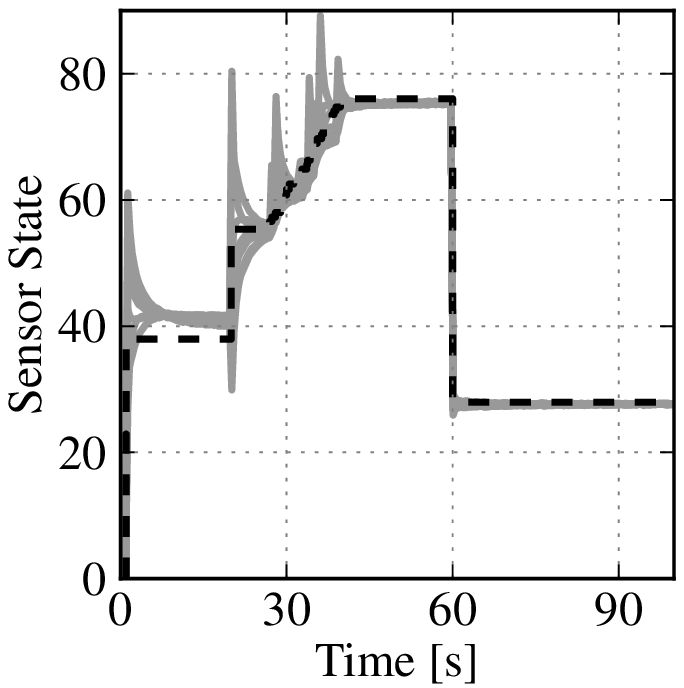}
 \label{fig:ring_30N}}
 \subfigure[Complete network]{
	\includegraphics[scale = 0.585]{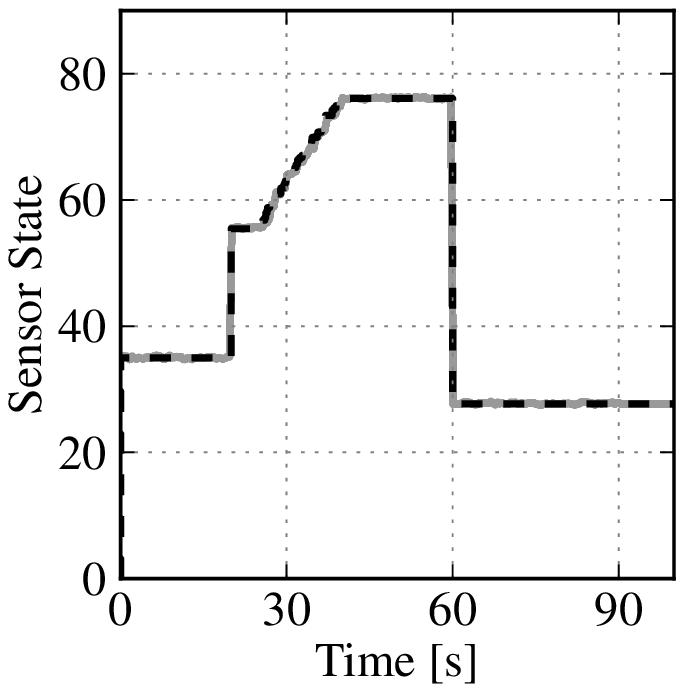}
 \label{fig:full_30N}
 }
\caption{{Sensors' state evolution when $M=30$ sensors are connected through (a) a ring and (b) a complete network topology. Sensors' measurements vary as follows: at time $t=0$, $20$, and $60$ (in s) the sensor measurements are independent random variables chosen uniformly within the intervals $[5, 30]$, $[50, 60]$ and $[25, 35]$, respectively. During the time interval between $[25,40]$, each sensor independently chooses a time instant to change its measurement, chosen uniformly within the interval $[75, 77]$.}}
\label{fig:simulFullRun}
\end{figure}

%%%%%%%%%%%%%%%%%%%%%%%%%%%%%%%%%%%%%
% COMPARISON %%%%%%%%%%%%%%%%%%%%%%%%%%%%%%%
\begin{figure}[t]
 \centering
 \subfigure[Deviation]{
	\includegraphics[scale = 0.585]{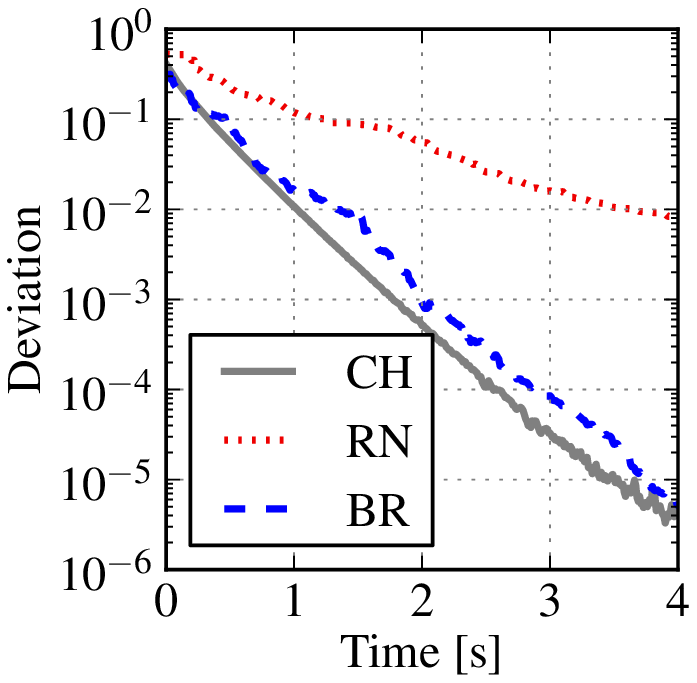}
 \label{fig:dev}
 }
% \hfil
 \subfigure[Mean squared error]{
	\includegraphics[scale = 0.585]{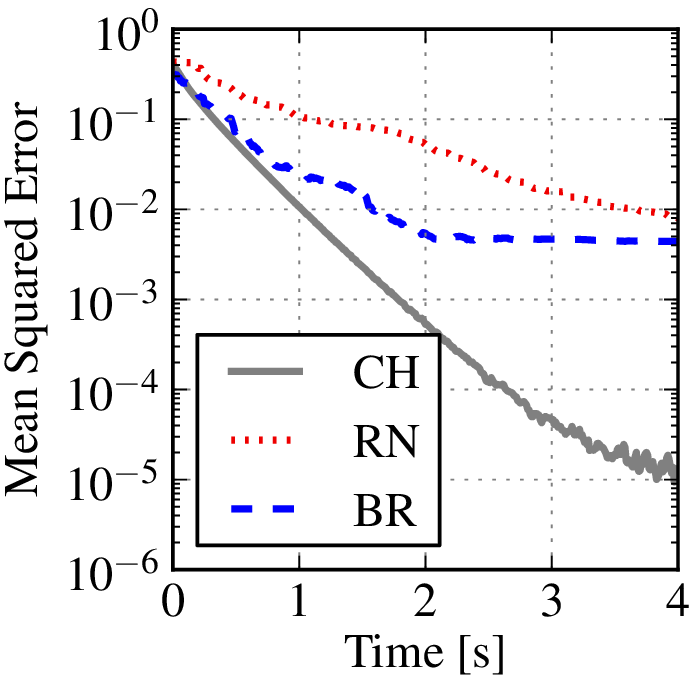}
 \label{fig:mse}
 }
\caption{{Performance comparisons among the chemical (CH) consensus algorithm, the broadcast (BR) and randomized (RN) gossip algorithms. Initial state and network setup are as in Fig~\ref{fig:100smallNet}.}}
\label{fig:compImpl}
\end{figure}

%%%%%%%%%%%%%%%%%%%%%%%%%%%%%%%%%%%%%%%%%%%%%%%%%%%%%%%%%%

{{Fig.~\ref{fig:forComparison_2} demonstrates the convergence of the sensor's state towards} the arithmetic mean for different network topologies (ring, complete, regular lattice, and small world networks), when the number of nodes is $M=25$. 
%Respectively, we study a ring () and a complete network (), and (Fig.~\ref{fig:forComparison_2}) a regular lattice network with interconnections to $k=3$ nearest neighbors and a small-world network with $3M =75$ links (see \cite{OlFa07} and references therein for more details on such networks). In all cases, the initial state is set to $z_i = i$ for $i=1,2,\ldots,M$. 
These results are in line with those in \cite{OlFa07}: compared to the other networks, small-world networks exhibit shorter convergence times, while still keeping the number of links reasonably low. This occurs thanks to the algebraic connectivity of the associated digraph $\mathcal G$.\footnote{The algebraic connectivity is the second smallest eigenvalue $\lambda_2$ of the associated digraph $\mathcal G$. The higher $\lambda_2$ is, the lower the convergence time is. Related to results in Fig.~\ref{fig:forComparison_2}, $\lambda_2$ is found to be $ 0.0314$, $25$, $0.8523$, and $2.0269$ for the ring, complete, regular lattice, and small-word networks, respectively.} To ease the comparison of the chemical algorithm's dynamics with those of the traditional model in \cite{OlFa07}, we also report in Fig.~\ref{fig:forComparison} the convergence time when $M=100$ sensors are connected through a regular lattice and a small-world network (refer to Fig.~4 in \cite{OlFa07}).}

%{Fig.~\ref{fig:forComparison} demonstrates the convergence in the same operating conditions of Fig.~\ref{fig:forComparison_2} for a network with $M=100$ nodes. As before, the convergence time of a regular lattice network is shown to be significantly larger than that of a small-word.} \tosee{The obtained simulation results are fully compliant with the numerical results presented in \cite{OlFa07}}.
%} 

{Table I reports the convergence times required by the artificial chemistry to reach a normalized mean squared error equal to $0.01$ for different numbers of nodes and different network topologies. The initial state is set as $z_1 = 60$ {and} $z_i=30$ with $i=2,3,\ldots,M$. } 
%In pa, we report the time $t_\epsilon$ required by the algorithm to reach a normalized mean squared error equal to $0.01$, \emph{vs.} the number of nodes, for ring, regular lattice, small world, and complete networks.
%%, i.e.,
%%\begin{equation}
%%\epsilon = \frac{\sum\nolimits_{i=1}^M c_{\mol{S}_i}(t_\epsilon) - z_{\text{avg}}}{z_{\text{avg}}}=0.01
%%\end{equation}
%%for different network topologies and number of sensors. 
%%Table \ref{tabTime} reports the results obtained for regular lattice networks with $k=3$ and small-world networks when $M=25, 50, 100, 250$ and $1000$. A number of links equal to $3M$ is used in small-world networks. In addition, the times required for ring and complete (fully connected) network topologies are also reported. 
%We refer to a regular lattice and a small-world network with $3M$ links. The initial states are set for simplicity equal to $z_1 = 30$ and $z_i = 15 $ for $i=2,3,\ldots,M$.
%It is worth observing that these ring and complete cases provide respectively a lower and upper bound to the number of connections links for a given number of nodes.} 
{From the results of Table I, we can see that the convergence time of a complete network decreases substantially as $M$ becomes larger {due to the exponential increase of the number of connected links.} The opposite happens for a ring network as, in this case, the information is exchanged in a serial manner and thus, the time required to exchange information among all nodes highly grows with increasing number of nodes.} {On the other hand, the convergence time remains constant for small-world networks.}

Fig.~\ref{fig:simulFullRun} illustrates sensors' state evolution when the network {is composed of} $M=30$ nodes and is characterized either by a ring or a complete topology. To validate the convergence of the algorithm in presence of measurement changes, the quantities $z_i$ {are randomly generated (at certain time instants) according to a uniform distribution} with given intervals (see Fig.~\ref{fig:compImpl} for details). From the experimental results in Fig.~\ref{fig:simulFullRun}, it follows that the convergence to each new value of the arithmetic mean is guaranteed for both network topologies. 
{This holds true only if the measured quantities vary sufficiently slow compared to the convergence time.}
As we can observe, due to the different algebraic connectivity, the convergence is achieved almost instantaneously for the complete network, whereas a longer time interval is required for the ring network. 

% Simul. figure: 	comparison chemical vs. broadcast vs. randomized gossip
{Comparisons are also made with the two following consensus gossip-based algorithms: the randomized (RN) solution proposed in \cite{BoGh06} and the broadcast (BR) one illustrated in \cite{AyYi09}. Both algorithms are simulated according to the asynchronous model described in \cite{BoGh06} and \cite{AyYi09}, whereby each node is assumed to have a clock that ticks independently according to a rate $\mu$ Poisson process. This corresponds to a single global clock whose ticking times form a Poisson process of rate $M\mu$ \cite{AyYi09}. In all subsequent simulations, we set an average of $\mu =2$ ticks per second in each node. When RN is used, at each tick, node $\nu_i$ randomly interacts with a single nearby sensor. On the other hand, node $\nu_i$ wirelessly broadcasts its current state value when BR is applied.} %interacts with nearby sensors a random selected nearby sensor interactions calibrated to trigger the broadcast transmission every 0.5s on average
 {Comparisons are made in terms of the normalized deviation of sensors' states from their average and of the mean squared error w.r.t. the average of their initial states (measured quantities $z_i$). 
}
{For this purpose, we consider the same operating conditions as in Fig.~\ref{fig:100smallNet}: a small world network with $M=100$ nodes and $3M = 300$ links in which the initial states are set to $z_i(0)=i$ for $i=1,2,\ldots,M$.} %\tosee{This means that, on average in the network, $2M\mu = 400$ packets per second are exchanged when RN is used, and $M\mu = 200$ packets per second are transmitted when BR is used}.
 %{As we can see, the proposed chemical (CH) consensus algorithm is able to exhibit similar performances to that of BR in terms of deviation, and better performances than both RN and BR in terms of mean squared error.}
 {The results of Fig. \ref{fig:dev} show that the convergence time of the proposed chemical (CH) algorithm is similar to that experienced with the BR algorithm. As seen, both largely outperform RN, which does not take advantage of the broadcast nature of the wireless medium. On the other hand, Fig. \ref{fig:mse} shows that the estimation accuracy of the CH algorithm is higher than that of the BR algorithm. This difference is due to the bias term that the BR algorithm introduces in the average estimation. For further details, the interested reader is referred to \cite{ShRa13}, in which a solution to overcome this problem is also discussed. However, this is achieved at the price of a higher convergence time and a more complex communication model that requires the transmission and the processing of a ``companion" variable, in addition to the node's state.} 
 %introduces a bias between the arithmetic and estimated mean (effect quantified in the limit shown by the mean squared error). 
 {%A possible solution to improve the estimation accuracy is to make use of knowledge of the number of neighbors. However, this comes at the price of a higher convergence time in terms of deviation, as illustrated in \cite{ShRa13}.
In summary, the results of Fig. \ref{fig:compImpl} show that the proposed CH algorithm allows one to achieve a good trade-off between convergence time and estimation accuracy.}

%%%%%%%%%%%%%%%%%%%%%%%%%%%%%%%%%%%%%%%%%%
%%%%%%%%%%%%%%%%%%%%%%%%%%%%%%%%%%%%%%%%%%
%%%%%%%%%%%%%%%%%%%%%%%%%%%%%%%%%%%%%%%%%%

\section{Some practical issues}\label{sec:Robustness}
{
In this section, {we discuss how to modify the chemistry-inspired dynamical system} of each sensor to account for some practical issues. 
}

%%%%%%%%%%%%%%%%%%%%%%%%%%%%%%%%%%%%%%%%%%
\subsection{Estimating the number of neighbors}
{Observe that the execution of $r_{i,D}$ in \eqref{eq:drainReaction} requires knowledge of $|\mathcal{N}_{i}|$, which is hardly available at each node, especially in those applications in which nodes appear and disappear over time.} 
{To address this issue,
we start rewriting \eqref{eq:remReaction} and \eqref{eq:drainReaction} as follows:
%\begin{subequations} \label{eq:neighEst0}
\begin{align} \label{eq:remReaction.1}
	r_{i,{B}^\prime}:&  \quad \mol{S}_{i}\stackrel{1}{\longrightarrow}  \sum_{j\, \in \, \mathcal{N}_i} \mol{S}_{j} + \mol{S}_{i}\\ \label{eq:drainReaction.1}
	r_{i, {D}^\prime}:& \quad \mol{S}_i  \stackrel{|\mathcal{N}_i|}{\longrightarrow} \emptyset. 
 \end{align}
% \end{subequations} 
The above operation has no effect on the dynamics of $\mathcal{AC}_i$ (same set of ODEs as in \eqref{eq:ODE_0}) and it is only used to make the system depend on $|\mathcal{N}_i|$ rather than $|\mathcal{N}_i| -1$.}\footnote{{This allows overcoming some implementation issues. First, computing the difference $|\mathcal N_i|-1$ would require an additional set of reactions (see for example the motif proposed in Section 10.3 of \cite{MeTh}). Second, using \eqref{eq:remReaction.1} and \eqref{eq:drainReaction.1} allows maintaining the sensor state also in a single-node network.}}
{To proceed further, we let each node $\nu_i$ define two molecular species $\mol{X}_i$ and $\mol{Y}_i$, with $\mol{X}_i$ characterized by a constant concentration equal to $\lambda$, i.e., {$c_{\mol{X}_i}(t) \equiv \lambda$}. Then, we define the following reactions:
\begin{subequations} \label{eq:neighEst}
\begin{align} 
\label{eq:remNR}
	r_{i,X}:&  \quad \mol{X}_i \stackrel{1}{\longrightarrow} \sum_{j \in \mathcal N_i}\mol{Y}_j + \mol{X}_i  \\  	\label{eq:locNR}
	r_{i,Y}:&  \quad \mol{Y}_i \stackrel{1}{\longrightarrow} \emptyset.
\end{align} 
\end{subequations} 
Using \eqref{eq:ODEs} and recalling that $c_{\mol{X}_i}(t) = \lambda$ yields
%\begin{equation}\label{eq:ODE_0_1_1}
%{\dot c_{{\mol{Y}_i}}} (t)=  {\sum\limits_{j \in {\mathcal{N}_i}}^{} {{c_{{\mol{X}_j}}}}(t)} -  {c_{\mol{Y}_i}}(t)
%\end{equation}
%or, equivalently, since $c_{\mol{X}_i}(t) = \lambda$ 
\begin{equation}\label{eq:ODE_0_1_11}
{\dot c_{{\mol{Y}_i}}} (t)=  \lambda |\mathcal{N}_i| -  {c_{\mol{Y}_i}}(t)
\end{equation}
from which it follows that at the equilibrium (i.e., ${\dot c_{{\mol{Y}_i}}} (t^\star) = 0$) the abundance of $\mol{Y}$-molecules at each node is $\lambda|\mathcal{N}_{i}|$:
\begin{equation}\label{eq:ODE_0_1_10}
{c_{\mol{Y}_i}}(t^\star) =\lambda |\mathcal{N}_i|.
\end{equation}
Then, replacing the draining reaction $r_{i, {D}^{\prime}}$ with
\begin{equation}\label{eq:drainReaction.2}
r_{i, {D}^{\prime\prime}}:\quad \mol{S}_i +\mol{Y}_i	  \stackrel{1/\lambda}{\longrightarrow} \mol{Y}_i	
\end{equation}
yields the following ODEs
\begin{equation}\label{eq:ODE_0_1_1_a}
{\dot c_{{\mol{S}_i}}} (t)=  {\sum\limits_{j \in {\mathcal{N}_i}}^{} {{c_{{\mol{S}_j}}}}(t)} -  \frac{1}{\lambda}{{c_{\mol{Y}_i}}(t){c_{\mol{S}_i}}(t)},\;\;c_{\mol{S}_i}(0)=z_i.
\end{equation}
from which, by assuming that the convergence time of \eqref{eq:remNR} and \eqref{eq:locNR} is smaller than that required by \eqref{eq:remReaction.1} and \eqref{eq:drainReaction.2} and thus by substituting \eqref{eq:ODE_0_1_10} in \eqref{eq:ODE_0_1_1_a}, we get the following result
\begin{equation}\label{eq:ODE_0_1_1}
{\dot c_{{\mol{S}_i}}} (t)=  {\sum\limits_{j \in {\mathcal{N}_i}}^{} {{c_{{\mol{S}_j}}}}(t)} -  |\mathcal N_i|{c_{\mol{S}_i}}(t),\;\;c_{\mol{S}_i}(0)=z_i.
\end{equation} 
Equation \eqref{eq:ODE_0_1_1} is in the same form as \eqref{eq:ODE_0} but has been obtained without knowing $|\mathcal{N}_{i}|$, simply by using the reactions in \eqref{eq:neighEst} and modifying the draining reaction as in \eqref{eq:drainReaction.2}.}

%\begin{equation}\label{eq:ODE_0_1_1}
%{\dot c_{{\mol{X}_i}}} (t)=  {\sum\limits_{j \in {\mathcal{N}_i}}^{} {{c_{{\mol{S}_j}}}}(t)} -  {{c_{\mol{Y}_i}}(t){c_{\mol{S}_i}}(t)},\;\;c_{\mol{S}_i}(0)=z_i.
%\end{equation}

{{\bf{Remark}}: Observe that the above results hold true only if \eqref{eq:remNR} and \eqref{eq:locNR} reach the equilibrium before \eqref{eq:remReaction.1} and \eqref{eq:drainReaction.2}. This is reasonable for low-mobility applications but can generally be achieved by properly setting the design parameter $\lambda$, which dictates the rate of execution of \eqref{eq:remNR}. The higher the $\lambda$-coefficient is, the faster the convergence is.} 

{{\bf{Remark}}: In those applications where $|\mathcal{N}_{i}|$ remains constant for a long time interval, its value can be easily estimated through artificial chemistry $\mathcal{AC}_i$ defined in \eqref{200} (with no need for additional reactions): During an initialization phase, the sensors' state should be maintained constant at the pre-defined value $\lambda$ (i.e., $c_{\mol{S}_i}(t) = \lambda$) and the reaction $r_{i,D}$ should not be locally executed. {In these cases}, the execution of $r_{i,B}$ would induce the production of \mol{S}-molecules in nearby sensors with an average rate $|\mathcal{N}_{i}|$ times bigger than the pre-defined value $\lambda$ (this easily follows recalling the broadcast nature of reaction $r_{i,B}$). Therefore, an estimate of $|\mathcal{N}_{i}|$ could be easily obtained by comparing the measured reception rate and the predefined one.} 

%%%%%%%%%%%%%%%%%%%%%%%%%%%%%%%%%
%%%%%%%%%%%%%%%%%%%%%%%%%%%%%%%%%

\subsection{Robustness to perturbations}
{A WSN must be robust to possible perturbations, such as measurement errors or sensors leaving (entering) the network in advance (at a later stage). To chemically address these issues, we let each node define a molecular species $\mol{Z}_i$ whose concentration is maintained constant and equal to the local measurement value, i.e., $c_{\mol{Z}_i}(t) = z_i$. Then, we introduce the two following reactions:
\begin{subequations} \label{eq:perturbReact}
\begin{align} 
		\label{eq:errIn}
	    	 r_{i,Z}:\quad&	 \mol{Z}_i \stackrel{\delta}{\longrightarrow} 	\mol{S}_i + \mol{Z}_i	\\
		 \label{eq:errDrain}
		 r_{i,A}:\quad &  \mol{S}_i \stackrel{\delta}{\longrightarrow} 	\emptyset 
\end{align}		
\end{subequations}  
with $\delta$ being a design parameter. The execution of $r_{i,Z}$ continuously feeds $\mol{S}_i$-species at a rate $\delta$-proportional to the local measurement $z_i$. At the same time, $\mol{S}_i$-species is drained with the same coefficient $\delta$ of proportionality through reaction $r_{i,A}$. As a result, sensors' state (the concentration of molecules $\mol{S}_i$) is continuously refreshed and if an error occurs, after a transient time, sensor's state goes back to the correct value. The higher the $\delta$-coefficient is, the faster the recovering is. As shown next by means of simulation results, this is however achieved at the price of a reduced estimation accuracy. 
%As shown next, the effect of this correcting subsystem is controllable through the $\delta$-coefficient. The higher the $\delta$-coefficient is, the faster the recovering is and the higher the differences between sensors estimates are. 

{Observe that the dynamics of (28)} are described by the following ODE:\footnote{{$\vet{z}$ represents the vector of initial states.}}
\begin{equation}\label{eq:ODE_0_28}
{\dot c_{{\mol{S}_i}}} (t)=  {\sum\limits_{j \in {\mathcal{N}_i}}^{} {{c_{{\mol{S}_j}}}}(t)} -  {|\mathcal N_i|}{c_{\mol{S}_i}}(t)  + \delta \left(z_i - {c_{\mol{S}_i}}(t) \right)
\end{equation}
whose matrix form is given by
%\begin{equation}\label{eq:ODE_0_29}
%{\dot c_{{\mol{S}_i}}} (t)=  {\sum\limits_{j \in {\mathcal{N}_i}}^{} {{c_{{\mol{S}_j}}}}(t)} -  \left({|\mathcal N_i|} + \delta\right){c_{\mol{S}_i}}(t)  + \delta {c_{\mol{S}_i}}(0)
%\end{equation}
\begin{equation}\label{eq:ODE_30}
{\dot {\mathbf{c}}_{{\mol{S}}}} (t)=  -{\mathbf{L}}{{\mathbf{c}}_{{\mol{S}}}} (t) + {\delta}\left(\mathbf{z}- {{\mathbf{c}}_{{\mol{S}}}} (t)\right)
\end{equation}
where $\mathbf{L}$ is the graph Laplacian of the network. Rewriting \eqref{eq:ODE_30} after taking the Laplace transform of both sides at a certain reference time $t=t^\prime$, we get
\begin{equation}\label{eq:ODE_31}
{ {\mathbf{C}}_{{\mol{S}}}} (s)=  {\mathbf{H}(s)}\left({{{\mathbf{c}}_{{\mol{S}}}} (t^\prime) + \delta \mathbf{z}}\right)
\end{equation}
where ${\mathbf{H}(s)}$ is the Laplacian transfer function given by
\begin{equation}\label{eq:H_32}
{\mathbf{H}(s)} = \left(s\mathbf{I}_M + {\mathbf{L}} + \delta\mathbf{I}_M\right)^{-1}
\end{equation}
with $\mathbf{I}_M$ being the identity matrix of order $M$. One can use ${\mathbf{H}(s)}$ to analytically evaluate how the $\delta$-coefficient must be chosen: a trade-off between the convergence time and estimation accuracy.

}

%%%%%%%%%%%%%%%%%%%%%%%%%%%%%%%%%%%%%
% RECOVER FROM ERROR FOR DIFFERENT DELTA %%%%%%%%%%%%%%%%%%%%%%%%%%%%%%%
\begin{figure}[t]
 \centering
 \subfigure[Sensors' state evolution]{
	\includegraphics[scale = 0.585]{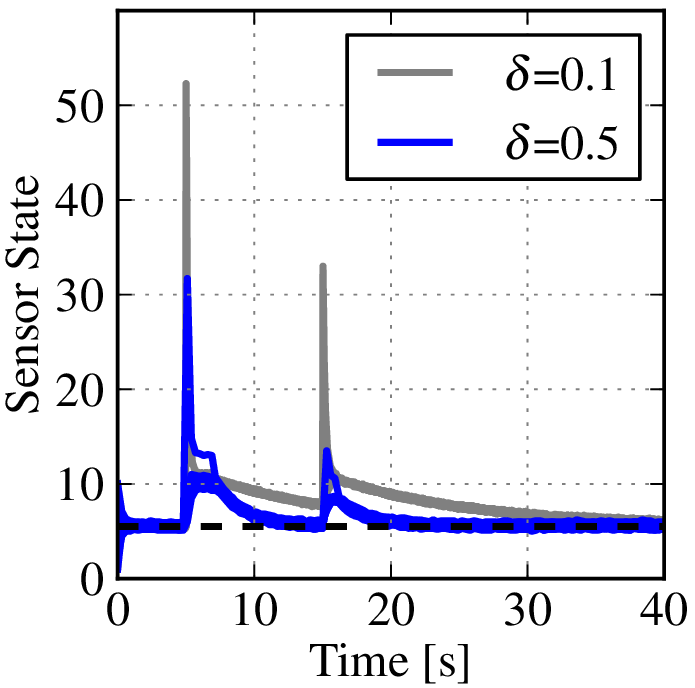}
 \label{fig:runErrFast}
 }
% \hfil
 \subfigure[Mean squared error]{
	\includegraphics[scale = 0.585]{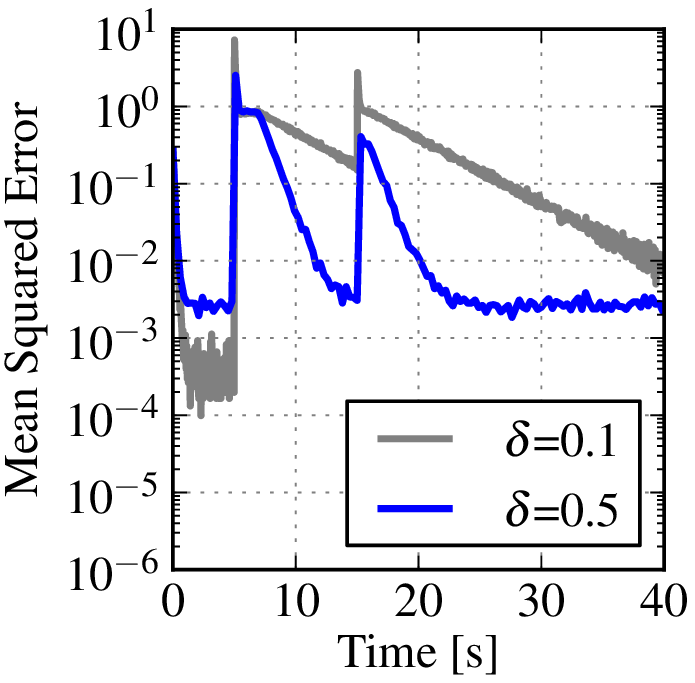}
 \label{fig:mseErrFast}
 }
\caption{{Effect of $\delta$ on the recovery from perturbations in a small world network with $M=10$ nodes, $3M= 30$ links. The initial state is $z_i=i$ for $i=1,2,\ldots,M$. At $t=5$, node $\nu_1$ introduces an error of 50\% for $2$ seconds. At $t=15$, node $\nu_4$ introduces an error of 30\% for $1$ second.}}
\label{fig:errFast}
\end{figure}

%%%%%%%%%%%%%%%%%%%%%%%%%%%%%%%%%%%%%
% COMPARISON ERROR %%%%%%%%%%%%%%%%%%%%%%%%%%%%%%%
\begin{figure}[t]
 \centering
 \subfigure[Nodes' state evolution]{
	\includegraphics[scale = 0.585]{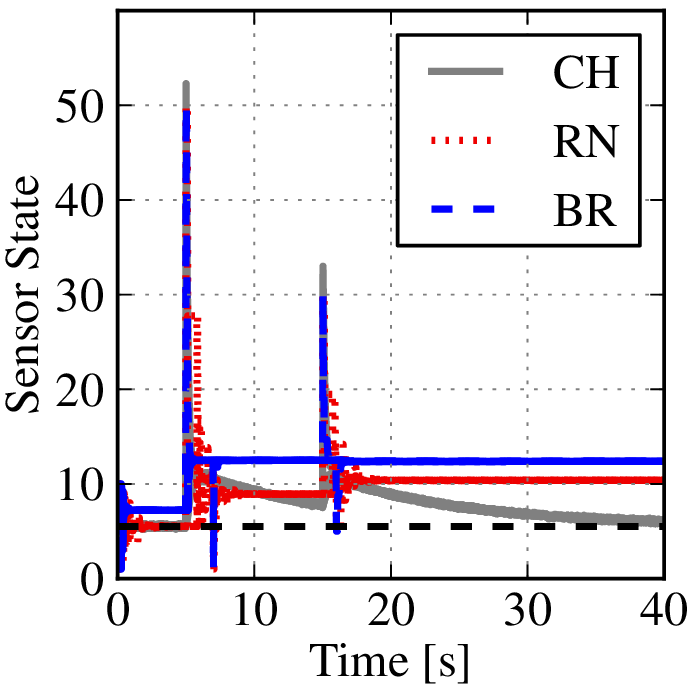}
 \label{fig:runErrComp}
 }
% \hfil
 \subfigure[Mean squared error]{
	\includegraphics[scale = 0.585]{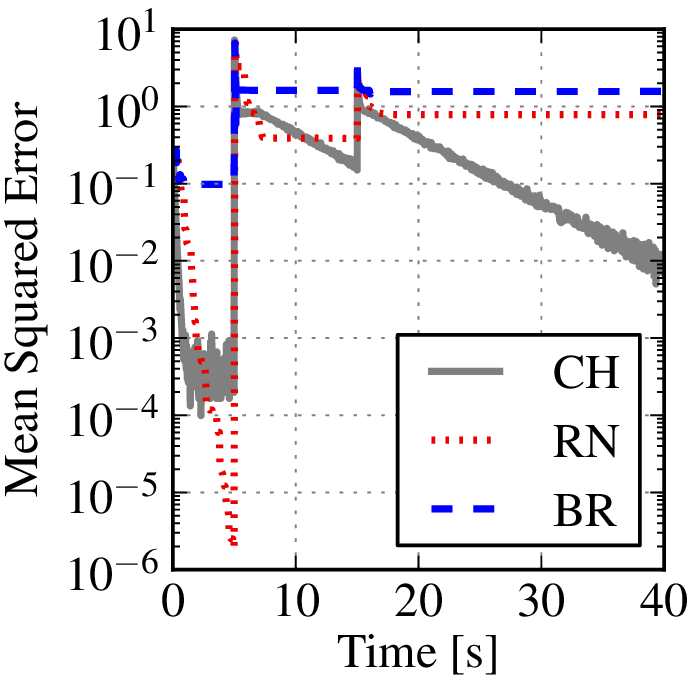}
 \label{fig:mseErrComp}
 }
\caption{{Performance comparisons among the chemical consensus (CH) algorithm and the broadcast (BR) and randomized (RN) gossip algorithms in presence of perturbations. At $t=5$, node $\nu_1$ introduces an error of 50\% for $2$ seconds. At $t=15$, node $\nu_4$ introduces an error of 30\% for $1$ second. Coefficient $\delta$ is set to $0.1$, and initial state and network setup are as described in Fig.~\ref{fig:errFast}. }}
\label{fig:errComp}
%\vspace{-0.05in}
\end{figure}

\subsection{Simulation results}

% Simul. figure: 	comparison chemical vs. broadcast vs. randomized gossip with error
{To account for the above mechanisms, the dynamical system of each node must simply be modified so as to operate according to the new artificial chemistry given by 
\begin{equation}\label{pippo_finale}
\mathcal{AC}_i^{\prime}=\{\mathcal M_i^{\prime}, \mathcal R_i^{\prime}, \mathcal A\} 
\end{equation}
with $\mathcal M_i^{\prime} = \mathcal{S}_i^{\prime} \cup \mathcal{S}_i^{(j)^{\prime}}$, $\mathcal{S}_i^{\prime} = \{\mol{S}_i,\mol{X}_i,\mol{Y}_i, \mol{Z}_i\}$, $\mathcal{S}_i^{(j)^{\prime}} = \{\mol{S}_j,\mol{Y}_j|\,j \in\mathcal{N}_i\}$ and 
\begin{align} \label{pluto_finale}
\mathcal R_i^{\prime} = \{r_{i,B^{\prime}},r_{i,D^{\prime\prime}}, r_{i,X}, r_{i,Y},r_{i,Z}, r_{i,A}\}.
\end{align}} 
% Simul. figure: 	comparison chemical vs. broadcast vs. randomized gossip when one or more nodes leave 
{We begin by assessing the impact of $\delta$. As mentioned before, the dimensioning of $\delta$ represents a trade-off between robustness to perturbations and estimation accuracy. In Fig.~\ref{fig:errFast}, we illustrate sensors' state evolution and mean squared error of CH algorithm when $\delta$ is either $0.1$ or $0.5$. 
%The investigated setting is a small world network with $M=10$ nodes, $3M= 30$ links and initial values equal to $z_i=i$ for $i=1,2,\ldots,M$.
In this experiment, certain nodes exhibit transient problems in sensing or transmitting, and thus introduce some perturbations in the chemical network.
%: at the time instant $t=5$, node $\nu_1$ introduces a measurement error for $2$ seconds while, at $t=15$, node $\nu_4$ introduces an error for $1$ second. 
The results of Fig.~\ref{fig:errFast} show that a higher value of $\delta$ allows a faster recovery from perturbations at the expense of a lower estimation accuracy. {As shown in Fig.~\ref{fig:mseErrFast}, the mean squared error during the first five seconds is less than $10^{-3}$ for $\delta=0.1$ whereas it is higher than $10^{-3}$ for $\delta=0.5$}. Simulation results (not shown for space limitations) show that $\delta = 0.1$ allows achieving a good tradeoff between the two conflicting requirements. For this reason, we set $\delta = 0.1$ in all subsequent simulations.\footnote{{Different values of parameter $\delta$ may be required in scenarios exhibiting different features (e.g., affected by perturbations of different intensity) or for different application constraints (e.g., a different estimation accuracy constraint).}}
} 

{Fig.~\ref{fig:errComp} illustrates performance comparisons among CH, BR, and RN algorithms, in the same perturbed scenario as in Fig.~\ref{fig:errFast}. The CH algorithm results to be resilient to measurement errors (perturbations) while RN and BR do not guarantee the achievement of average consensus. A simple (but inefficient) solution to make RN and BR recover from errors would be that of including a mechanism that automatically switches off all sensors and lets them run with the new measurements. However, this should be done whenever a perturbation occurs.}

% Simul. figure: 	comparison chemical vs. broadcast vs. randomized gossip when one or more nodes leave 
%{Fig.~\ref{fig:offComp} illustrates the performance of CH using the same network model of Fig.~\ref{fig:errComp} but assuming that nodes $\nu_1$ and $\nu_3$ suddenly disappear at time $t=10$ and $t=15$, respectively. As seen, the correction mechanism introduced in CH lets the algorithm to track the variation of the average value while RN and BR fail.
%} 

\begin{figure}[t]
 \centering
 \subfigure[Nodes' state evolution]{
	\includegraphics[scale = 0.585]{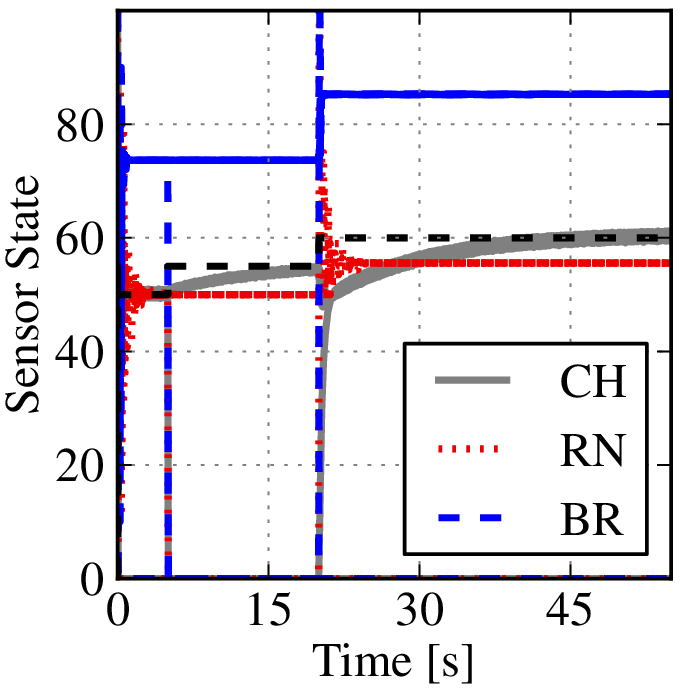}
 \label{fig:runOnOffComp}
 }
% \hfil
 \subfigure[Mean squared error]{
	\includegraphics[scale = 0.585]{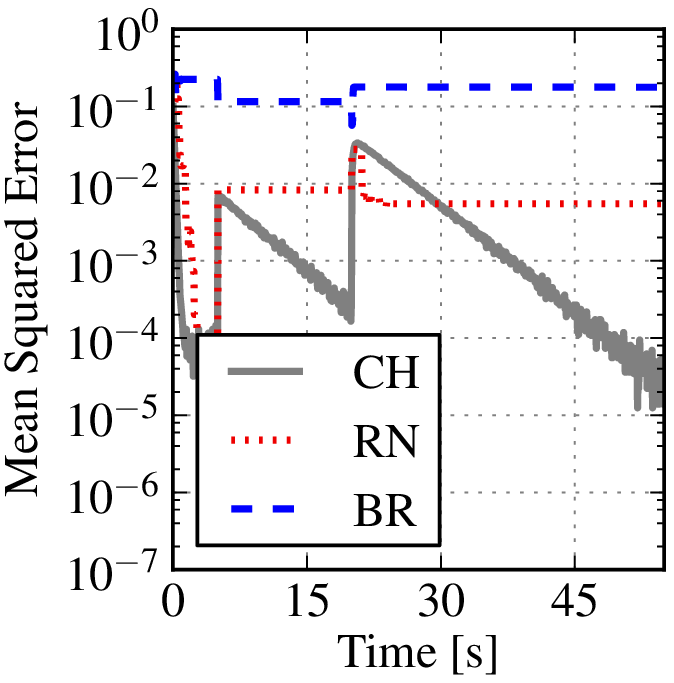}
 \label{fig:mseOnOffComp}
 }
\caption{{Performance comparisons among the chemical consensus (CH) algorithm and the broadcast (BR) and randomized (RN) gossip algorithms in presence of perturbations. The initial state is $z_i=10i$ for $i=1,2,\ldots,M$. At $t=5$, node $\nu_1$ disappears whereas node $\nu_{10}$ switches on at $t=20$. Coefficient $\delta$ is set to $0.1$ and the network setup is as described in Fig.~\ref{fig:errFast}.}}
\label{fig:onoffComp}
\end{figure}

% Simul. figure: 	comparison chemical vs. broadcast vs. randomized gossip when one or more nodes leave 
{Fig.~\ref{fig:onoffComp} illustrates the performance of CH, RN, and BR in the operating conditions described in Fig.~\ref{fig:errFast}. This time, node $\nu_1$ suddenly disappears at time $t=5$ whereas node $\nu_{10}$ only appears at $t=20$. As we can see, the correction mechanism introduced in CH lets nodes track the variations induced in the average value by the intermitting communications, while RN and BR fail.
}

%%%%%%%%%%%%%%%%%%%%%%%%%%%%%%%%%%%%%%
%% COMPARISON OFF %%%%%%%%%%%%%%%%%%%%%%%%%%%%%%%
%\begin{figure}[t]
% \centering
% \subfigure[Nodes' state evolution]{
%	\includegraphics[scale = 0.585]{figs/py/bigFigure/off_comparison/run}
% \label{fig:runOffComp}
% }
%% \hfil
% \subfigure[mean squared error]{
%	\includegraphics[scale = 0.585]{figs/py/bigFigure/off_comparison/off_meansquarederr}
% \label{fig:mseOffComp}
% }
%\caption{Comparison between chemical consensus (``CH"), broadcast gossip~(``BR''),  and randomized gossip algorithms (``RN") in case of sensor faults: All sensors $\nu_i$ were initialized as $z_i(0)=i$; at time 10s $\nu_1$ disappeared; at time 15s also switched off. $\delta=0.1$.}
%\label{fig:offComp}
%\end{figure}

%%%%%%%%%%%%%%%%%%%%%%%%%%%%%%%%%%%%%
% COMPARISON ON-OFF %%%%%%%%%%%%%%%%%%%%%%%%%%%%%%%

%% Simul. figure: 	perturbation and self healing for different nets.
%The results of Fig.~\ref{fig:simErr} show the effectiveness of the proposed solution against measurement errors{. In} particular, we report the time required by the network to recover from impulsive measurement errors. The latter are assumed to be equal to 30\% of local measurements. The number of sensors is fixed to $10$ since errors are expected to have high influence in networks with a relatively small number of nodes. From the results of Fig.~\ref{fig:simErr}, it follows that {nodes recover from errors and converge back to the arithmetic mean in a finite time.}%, especially in ring networks. 
%
%% Simul. figure: 	effect of delays and correcting mechanism.
%In Fig.~\ref{fig:SimDel}, we deal with communication delays and consider either a complete or a ring network with $30$ sensors. Communication delays were assumed to be random variables uniformly distributed within the interval $\mathcal{U}[0,100]$ms. Results refer to a system in which $\delta$ was set to $0.01$. %We made also comparisons for a system in which no correcting mechanism was included ($\delta =0$) and a system with $\delta =0.1$. We experienced that higher values of $\delta $ guarantee a shorter convergence time but result in large deviations from the consensus.

%%%%%%%%%%%%%%%%%%%%%%%%%%%%%%%
\section{Hardware implementation and experimental results}\label{sec:HWimpl}
\begin{figure}[t]
 %\vspace{-0.15in}
 \centering
\centerline{\includegraphics[scale =0.27]{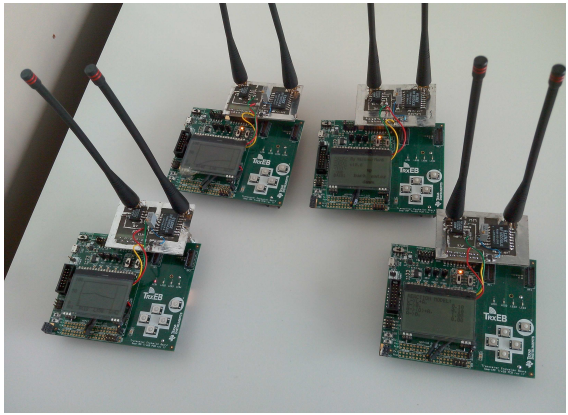}}
\caption{Experiment testbed consisting of four sensors where the dynamical system is implemented in the embedded processor TI-MSP430F5438A, and the radio transceiver through transmitter TXM-433-LR, receiver RXM-433-LR, and comparator MAX-921. Sensors are implemented on TI-CC1120 evaluation board which has also an LCD display for a practical setup and tracking of run-time computation.}
\label{fig:photo}
\end{figure}

{In this section, we report on the experimental results obtained with a four-node hardware implementation (see Fig.~\ref{fig:photo}), operating according to the artificial chemistries defined in \eqref{pippo_finale} and \eqref{pluto_finale}. To our knowledge, this is the first time that a chemistry-inspired algorithm is built in a hardware testbed and validated under real-word conditions, where nodes exchange their data in an asynchronous manner with no need for admission control. %Therefore, we consider all this as one of the major contributions of this work.
}

% PHOTO %%%%%%%%%%%%%%%%%%%%%%%%%%%%%%%%

%%%%%%%%%%%%%%%%%%%%%%%%%%%%%%%%%%%%%%%%% 
\subsection{Hardware implementation}

The dynamical system, operating according to artificial chemistry specifications, is implemented into an embedded processor TI-MSP430F5438A, which is also used for data acquirement and conversion. The radio interface (transceiver) is developed with a simple low-cost circuit using the TXM-433-LR integrated chip for transmitting to and the RXM-433-LR integrated chip for receiving from nearby nodes. {To limit the complexity of each node, we let the sensor interactions occur in a simple manner. Specifically, we assume that a pulse $g_\tau(t)$ of duration $\tau$ is sent over the channel whenever the remote reaction $r_{i,{B^{\prime}}}$ in \eqref{pluto_finale} is executed (in the dynamical system). On the other hand, the production of an $\mol{S}$-molecule is induced whenever a pulse $g_\tau(t)$ is received from nearby sensors. According to the law of mass action in \eqref{eq:LoMA}, the average rate of occurrence of $r_{i,{B^{\prime}}}$ is proportional to the concentration of $\mol{S}$-molecules (sensor's state). This means that the spatially distributed nodes interact by means of  pulses whose transmission rate encodes nodes' state. Following the same line of reasoning, we let a pulse $g_{\tilde \tau}(t)$ of duration~$\tilde \tau \ne \tau$ be sent whenever the remote reaction $r_{i,X}$ is executed, and a $\mol{Y}$-molecule be produced whenever $g_{\tilde \tau}(t)$ is received.} 

{
{\bf{Remark}} To ease understanding, consider a simple network only composed by $M=2$ nodes. Without loss of generality, we concentrate on the first node and explain how sensor interactions occur. The node starts transmitting pulses $g_\tau(t)$ at a rate equal to the initial concentration of its local species (value equal to the measured quantity, i.e., $c_{S_1}(0)=z_1$). At the same time, $c_{S_1}(t)$ is continuously modified on the basis of the number of $g_\tau(t)$-pulses that are received from node 2. This happens at a rate equal to $c_{S_2}(t)$. Internally, the current state $c_{S_1}(t)$ is also continuously decreased at a rate $c_{S_1}(t)$, according to reaction $r_{1,{D^{\prime}}}$ in \eqref{eq:drainReaction.1}, and modified proportionally to the $\delta$-parameter, according to the reactions reported in \eqref{eq:perturbReact}. To continuously estimate the number of neighbors, node 1 has also to transmit $g_{\tilde \tau}(t)$-pulses at a constant rate $\lambda$ and to increase $c_{Y_1}(t)$ whenever a pulse $g_{\tilde \tau}(t)$ is received from node 2. 
}

{{\bf{Remark}}: Observe that the pulse rate depends on the sensor's state. However, this does not mean that if $z_i = 10^9$ then $10^9$ pulses $g_\tau(t)$ must be transmitted by node $\nu_i$. Indeed, it is important to decide how the values are encoded or, in other words, how a molecule quantity has to be interpreted. Assume for example that the algorithm has to measure the average temperature in a sensor network. Then, in order to limit the pulse rate, one has to properly associate the right quantity to a single molecule instance (degrees celsius, Kelvin, Fahrenheit). This allows controlling the maximum transmission rate at the price of a reduced accuracy of the computation
(accuracy in the estimation).}

{{\bf{Remark}}: A possible drawback of the above implementation is that no countermeasures are taken against interferences that might arise in WSNs, when the signals transmitted by multiple nodes collide at a given receiving node. Although a judicious design of the system parameters (maximum value of concentrations, duration of the pulse and so forth) could reduce the occurrence of collisions, more advanced multiple access protocols are required to effectively counteract the above issue, thus increasing the complexity of each sensor. For this reason, we have decided not to take countermeasures against interferences. {This choice has also been motivated by the observation that chemical systems usually exhibit strong robustness to perturbations thanks to the mass-action kinetics governing their interaction mechanisms (the interested reader is referred to \cite{ShAl09} for a recent work in the context of sensitivity and robustness of chemical reaction systems). Therefore, a chemistry-inspired algorithm is likely to be robust against the unreliable conditions of WSNs.\footnote{{The robustness is a direct consequence of the fact that the information exchange is encoded into a rate rather than in one or few information packets, and thus any corruption of one or few of these transmissions does not affect significantly the system. Another reason for the inherent robustness is that mass action kinetics often induces low-pass filtering behaviors and transfer functions exhibiting negative real-part poles.}} {This has indeed been confirmed in our experiments although only four-node WSNs were tested}.}%{Although limited in scope, our experimental results corroborate this belief}.}% This has been in fact validated with our experimental results.

\subsection{Experimental results}

The transmitter output power is $3$ dBm on average whereas the receiver sensitivity is $-120$ dBm with a dynamic range of $80$ dB. {Each sensor is roughly characterized by a total power consumption of $36$ mW approximately distributed as follows: $13$ mW consumed by the embedded processor, $17$ mW by the receiver, and $6$ mW by the transmitter. The nodes are placed approximately 7 meters apart from each others, in an indoor environment.}
The transmission takes place over the free Industrial Scientific Medical radio band $433.05 \div 434.79$ MHz around the carrier frequency of $433.92$ MHz. We set $\tau=200$ $\mu s$ and $\tilde \tau=100$ $\mu s$, and directly control the quantities $z_i$ (sensor measurements) during the experiment, as shown in Table \ref{tbl:localData}. 
%Specifically, at the time instant $t=0$ (expressed in seconds) we set $z_1 = 50$ and $z_2=z_3=z_4=0$. At $t = 9.8$, $z_2$ is changed to $50$ whereas $z_3$ is set to $20$ at $t= 19.9$. At $t = 53$, we start decreasing $z_2$ to reach $z_2=0$ at $t = 57$. Finally, we let node $\nu_2$ leave the network at $t = 63$. 
In order to test the robustness of the network against external interferences, we also intentionally let a radio signal interfere with the information exchange within the time interval $[22, 33]$.

% REAL_ESP TAB %%%%%%%%%%%%%%%%%%%%%%%
\begin{table}[t]
  \caption{Local data $z_i$ detected by node $\nu_i$ at time $t$ during the experiment whose results are plotted in Fig.\ref{fig:realExpPlot}.}
  \label{tbl:localData}
  \centering
  \begin{tabular}{| c || c | c | c | c | c |} 
  \hline   
 Time instant $t $ [seconds] & 0  & 9.8     & 19.9 & 53 $\rightarrow$ 57			& 63\\\hline
 \hline  
 Value of $z_1$& 50& 50     & 50 & 50			&50\\\hline
 Value of $z_2$&  0 & 	 50 & 50 & $\rightarrow$ 0& \slash \\\hline
 Value of $z_3$& 0  & 	 0   & 20 & 20			& 0\\\hline
 Value of $z_4$&  0 & 	 0   & 0   & 0			& 0\\
   \hline 
  \end{tabular}
\end{table} 
%%%%%%%%%%%%%%%%%%%%%%%%%%%%%%%%%%%%%%%%%%%%%%
% REAL_ESP PLOTS  %%%%%%%%%%%%%%%%%%%%%%%%%%%%%%%%
\begin{figure}
\centering
\includegraphics[scale=0.63]{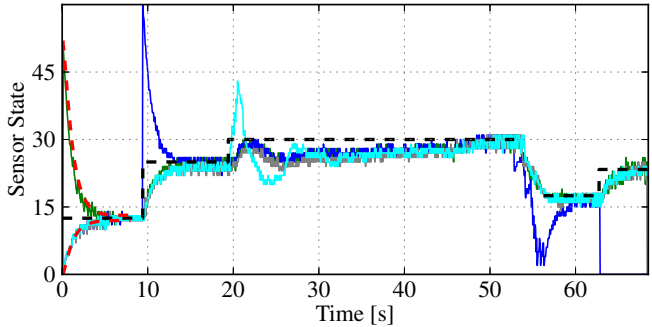}
\caption{Experiment results (4 colored, continuous lines) and analytical predictions obtained from \eqref{eq:H_32} (2 red, dashed lines -- within the time interval going from $t=0$s to $t=9$), and arithmetic mean (1 black, dashed line). The $\delta$ coefficient is set to $0.1$. 
}
\label{fig:realExpPlot}
\end{figure}
%
%Experiment succeeds: Figure description
%Measurements
Fig.~\ref{fig:realExpPlot} illustrates sensors' state  evolution (taken at a sampling rate of $10$ ms) of each node $\nu_i$ under the above operating conditions. The green, blue, cyan and grey continuous-lines refer to sensors' state  evolution of node $\nu_1$, $\nu_2$, $\nu_3$ and $\nu_4$, respectively.
%Arithmetic means
{The black dashed-line represents the arithmetic mean as obtained through \eqref{eq:scope} while the red dashed-lines represent the analytical responses of sensor local states as obtained using the Laplacian transfer function in \eqref{eq:H_32}. For illustration purposes, only the first $9$ seconds of the analytical responses are shown. As we can see, the experimental results are in line with the analytical ones and show that the nodes are able to correctly converge to the desired value after a ``short" transient time, regardless of measurement variations and interferences. In particular, it turns out that the effect of the interfering signal (from $t=22$ to $t=33$) is that of temporarily making the sensors underestimate the average. Once removed, each sensor converges to the desired value in approximately $20$ seconds.}

\section{Discussions and open issues}\label{sec:Discussion}
{This paper is basically divided into two parts:}

\subsubsection{Artificial chemistry for consensus in WSNs}
{We first show analytically and by means of simulations that simple interaction mechanisms inspired by chemical systems can provide the basic tools for achieving consensus in WSNs. However, some issues are still open.}

% Discretization
{In this work, we do not cover in details the discretization aspect (studied for example in \cite{CaFa10} and summarized in \cite{DiKa10}), which may affect the final nodes' average-estimate. We limit to observe here that by calibrating the amount of produced molecules per sensed-quantity unit, designers can regulate the precision of the nodes' estimates and decide the amount of transmissions.\footnote{In the simulations, we have coped anyway with the effect of discretization: we have had to use a number of molecules per unit quantity so as to obtain acceptable estimation accuracy.} 
%Reduce the number of transmissions
Please observe further that another way to reduce the computations/transmissions, and thus the {resource consumption}, consists in slowing down the virtual time characterizing the chemical model: transmissions decrease in number, this time, at the cost of a proportionally-slower adaptation (in our simple implementation, lower rates mean lower probability of collisions/interferences).}

%no channel models
{In this work, we have {considered a very simple communication model and referred not to} specific channel models and communication technologies. However, we have tested by means of simulations the ability of a chemistry-based system to recover from perturbations. In fact, such perturbations may represent the generalized effect of fading (transient under-valued state of one or more sensors), multipath (transient over-valued state of one or more sensors), or non-reliable links (variation of the number of nodes and different participation times).  
}

%balanced digraph only
{This work has focused on balanced digraphs only. In the presence of unbalanced graphs, the theoretical and experimental results illustrated in this paper are no longer valid. Indeed in this case, sensors \emph{converge} to a common value that differs from the average of the local measurements.\footnote{For unbalanced graphs, this work does not represent a solution to constrained consensus problems but it is still a solution to unconstrained consensus problems, according to definitions in \cite{OlFa07}.} This result occurs because, at the equilibrium, molecules are still evenly distributed among all participating sensors but {are ``diffused" with different proportions}, depending on the ratios between the number of receiving neighbors and that of transmitting ones. We are currently working on the development of solutions able to take into account such different weights while still using chemical mechanisms of similar complexity to those adopted in this paper. }

\subsubsection{A simple implementation of artificial chemistry in WSNs}
{The second part of the paper shows that the artificial chemistry can work under real-world conditions, and that the above framework can be used directly to develop a hardware implementation. To our knowledge, this is the first time we go beyond the theoretical treatment of chemical algorithms, and prove the applicability (at the lowest networking layers) of the chemical approach in a real sensor network. 
Although the experimental results are quite promising even under real-world conditions, this work should not be seen as a finalized, ready-to-use commercial product for nowadays markets. We believe that further research in the implementation context may bring significant improvements in terms of robustness and speed.
}

%%%%%%%%%%%%%%%%%%%%%%%%%%%%%%%%%%%%%%%%%%%%%%%
%%%%%%%%%%%%%%%%%%%%%%%%%%%%%%%%%%%%%%%%%%%%%%%
\section{Conclusions}
\label{sec:Conclusion}

In this paper, we have made use of distributed artificial chemistry to derive and analyze a set of interaction rules that allows achieving consensus in WSNs in a distributed manner, with no need for any synchronism and admission control mechanism. The proposed solution has been first validated and compared with other solutions, by means of experimental results, and then tested under real-world conditions using a four-node hardware implementation. The numerical and experimental results show that the use of artificial chemistry for deriving, analyzing, and implementing communication protocols is not merely an intellectual exercise but an {alternative} approach, which may pave the way for the development of robust solutions, able to cope with the uncertainties of WSNs. %To our knowledge, this is the first time that such a thorough treatment is carried out and we hope it may serve as an incentive to the research community for further investigations.
 
\section*{Appendix A: Reaction algorithm}
{The reaction algorithm $\mathcal A$ is reported below.}
{
\begin{enumerate}
	\item Initialize: 
	\begin{enumerate}
			\item set the initial amount of molecules;
			\item calculate the value $v_r$ according to \eqref{eq:LoMA} $\forall r\in\mathcal{R}$; 
			\item set a putative reaction-execution time $t_r = 1/v_r$ $\forall r\in\mathcal{R}$;
			\item store $t_r$ values in an indexed priority queue (first stored element has the next reaction time). 				\end{enumerate}		
	\item Let $r_\mu$ be the reaction whose putative reaction time, $t_\mu$, is least. 
	\item Wait as long as $t<t_\mu$.
	\item Change the number of molecules to reflect execution of reaction $r_\mu$. 
	\item Update all those reactions, $r_\alpha$, that depend on the executed reaction $r_\mu$:
	\begin{enumerate}
			\item  temporarily store the old value $v_\alpha^{old}=v_\alpha$; 
			\item  calculate the new value $v_\alpha$, according to \eqref{eq:LoMA}; 
			\item if $r_ \alpha \neq r_\mu$, scale the reaction execution time as $t_\alpha = (v_\alpha^{old}/v_\alpha)(t_\alpha-t) + t$; \\
			else if $r_ \alpha = r_\mu$, set the reaction execution time $t_\alpha = 1/v_\alpha +t $; 
			\item store the calculated reaction execution time $t_\alpha$ in the indexed priority queue.
	\end{enumerate}
	\item Go to Step 2.
	\end{enumerate}
	$*$~~The variable $t$ reflects the current time.\\
	$**$~Molecular species represent mere counters. Step~4 implies to decrement all those counters related to the reagent-species (species appearing in \eqref{eq:reaction} on the left-hand side of the arrow) and to increment all those counters related to the product-species (species appearing in \eqref{eq:reaction} on the right-hand side of the arrow).\\
	$***$~We consider deterministic inter-reaction times.
}

\section*{Appendix B: Convergence and stability}
%{In the next, we prove that the above set of ODEs admits a single fixed point solution given by \eqref{eq:scope} and demonstrates also that it is asymptotically stable. We do that not resorting to classical methods used in gossip context but rather exploiting results gained thanks to the import of chemical theories.}
{We start observing that the dynamics of the reaction network emerging from \eqref{eq:remReaction} and \eqref{eq:drainReaction} at each sensor node $\nu_i\in\mathcal V$ are equivalent to those obtained by deploying the following set of reactions at each node $\nu_i$:
\begin{equation}\label{eq:A.1}
	\mol{S}_i \longrightarrow \mol{S}_j \;\; | \;\; \nu_j \in \mathcal N_i.
\end{equation}
}
To proceed further, we need to briefly introduce the terms ``complexes'' and ``weakly reversible''. Complexes are those multisets of species that appear on the left- and the right-hand side of a reaction,~\cite{HoJa72}. A chemical reaction network is weakly reversible if for every reaction leading from complex $C_i$ to complex $C_j$, there is also a chain of reactions leading from $C_j$ back to $C_i$. 
{According to this definition, the reaction network arising from \eqref{eq:A.1} is weakly reversible. To see how this comes about, observe that all the species of the reaction network are complexes. Moreover, they do correspond also to the graph {vertices }of the communication network. This means that in strongly connected graphs for every reaction leading from complex $C_i$ to complex $C_j$ there exists a chain of reactions leading from $C_j$ to $C_i$. In addition, the emergent reaction network is closed (the amount of molecules within the system remains constant).}

{We now introduce the ``deficiency'' defined as  
\begin{equation}\label{deficiency}
	\gamma= |\mathcal C| - l - \textrm{rank}(\mat{U})
\end{equation}
where $\mathcal C$ denotes the set of complexes, $l$ is the number of linkage classes (i.e., the number of connected {subgraphs} in the graph of complexes), and $\textrm{rank}(\mat{U})$ denotes the rank of the stoichiometric matrix $\mat{U}$. From \eqref{eq:A.1}, it follows that $|\mathcal C| = |\mathcal{V}|$ and it can easily be proven that $ l = 1$ since each complex is connected directly or indirectly to any other complex constituting the whole digraph. Moreover, a strongly connected chemical reaction network, where every chemical species appears in precisely one complex, has a stoichiometric matrix with rank equal to $\textrm{rank}(\mat{U})= |\mathcal{V}|-1$~\cite{LeAn05}.} Collecting all the above facts together, we have that the deficiency of the reaction network associated to \eqref{eq:A.1} is {zero}. 

{According to the Deficiency Zero Theorem~\cite{MaHo74}, if the reaction network is weakly reversible and has {a null} deficiency value then it has a single, asymptotically-stable fixed point. Setting $\dot c_{{\mol{S}_i}} (t)= 0$ and studying the equilibrium solution, it follows that the fixed point is defined by the right-hand side of \eqref{eq:scope}. This proves that the proposed chemical algorithm converges to the average of initial measurements.}

\bibliographystyle{IEEEtran}
\bibliography{references}

\begin{thebibliography}{10}
\providecommand{\url}[1]{#1}
\csname url@samestyle\endcsname
\providecommand{\newblock}{\relax}
\providecommand{\bibinfo}[2]{#2}
\providecommand{\BIBentrySTDinterwordspacing}{\spaceskip=0pt\relax}
\providecommand{\BIBentryALTinterwordstretchfactor}{4}
\providecommand{\BIBentryALTinterwordspacing}{\spaceskip=\fontdimen2\font plus
\BIBentryALTinterwordstretchfactor\fontdimen3\font minus
  \fontdimen4\font\relax}
\providecommand{\BIBforeignlanguage}[2]{{%
\expandafter\ifx\csname l@#1\endcsname\relax
\typeout{** WARNING: IEEEtran.bst: No hyphenation pattern has been}%
\typeout{** loaded for the language `#1'. Using the pattern for}%
\typeout{** the default language instead.}%
\else
\language=\csname l@#1\endcsname
\fi
#2}}
\providecommand{\BIBdecl}{\relax}
\BIBdecl

\bibitem{OlFa07}
R.~Olfati-Saber, J.~Fax, and R.~Murray, ``Consensus and cooperation in
  networked multi-agent systems,'' \emph{Proceedings of the {IEEE}}, vol.~95,
  no.~1, pp. 215--233, Jan 2007.

\bibitem{AyYi09}
T.~C. Aysal, M.~E. Yildiz, A.~D. Sarwate, and A.~Scaglione, ``Broadcast gossip
  algorithms for consensus,'' \emph{IEEE Transactions on Signal Processing},
  vol.~57, no.~7, pp. 2748--2761, Jul 2009.

\bibitem{FrGi11}
M.~Franceschelli, A.~Giua, and C.~Seatzu, ``Distributed averaging in sensor
  networks based on broadcast gossip algorithms,'' \emph{IEEE Sensors J.},
  vol.~11, no.~3, pp. 808--817, Mar 2011.

\bibitem{DiKa10}
A.~G. Dimakis, S.~Kar, J.~M.~F. Moura, M.~G. Rabbat, and A.~Scaglione, ``Gossip
  algorithms for distributed signal processing,'' \emph{Proceedings of the
  IEEE}, vol.~98, no.~11, pp. 1847--1864, Nov 2010.

\bibitem{Barbarossa2007}
S.~Barbarossa and G.~Scutari, ``Bio-inspired sensor network design,''
  \emph{IEEE Signal Processing Magazine}, vol.~24, no.~3, pp. 26--35, May 2007.

\bibitem{Charalambous2010}
C.~Charalambous and S.~Cui, ``A biologically inspired networking model for
  wireless sensor networks,'' \emph{IEEE Network}, vol.~24, no.~3, pp. 6--13,
  Jun 2010.

\bibitem{WuGu07}
Z.~Wu, Z.~Guan, Z.~Wu, and T.~Li, ``Consensus based formation control and
  trajectory tracing of multi-agent robot systems,'' \emph{J. of Intelligent
  and Robotic Systems}, vol.~48, no.~3, pp. 397--410, Mar 2007.

\bibitem{BoGh06}
S.~P. Boyd, A.~Ghosh, B.~Prabhakar, and D.~Shah, ``Randomized gossip
  algorithms,'' \emph{IEEE Transactions on Information Theory}, vol.~52, no.~6,
  pp. 2508--2530, Jun 2006.

\bibitem{SiSp07}
O.~Simeone and U.~Spagnolini, ``Distributed time synchronization in wireless
  sensor networks with coupled discrete-time oscillators,'' \emph{EURASIP J. on
  Wireless Communications and Networking}, pp. 1--13, 2007.

\bibitem{CoGi06}
I.~Councill, L.~Giles, and P.~Teregowda, ``Consensus propagation,'' \emph{IEEE
  Transactions on Information Theory}, vol.~52, no.~11, pp. 4753--4766, Nov
  2006.

\bibitem{AvEl11}
K.~Avrachenkov, M.~E. Chamie, and G.~Neglia, ``A local average consensus
  algorithm for wireless sensor networks,'' in \emph{Proc. of IEEE Conf. on
  Distributed Computing in Sensor Systems (DCOSS)}, Barcelona, Spain, Jun 2011,
  pp. 1--6.

\bibitem{SiSp08}
O.~Simeone, U.~Spagnolini, G.~Scutari, and Y.~Bar-Ness, ``Physical-layer
  distributed synchronization in wireless networks and applications,''
  \emph{Physical Communication}, vol.~1, no.~1, pp. 67--83, Mar 2008.

\bibitem{NoBa11}
M.~Nokleby, W.~U. Bajwa, R.~Calderbank, and B.~Aazhang, ``Gossiping in groups:
  Distributed averaging over the wireless medium,'' in \emph{Proc. of the 49th
  Annual Allerton Conf. on Communication, Control, and Computing}, Monticello
  (IL), USA., Sep 2011, pp. 1242--1249.

\bibitem{NaDi11}
B.~Nazer, A.~G. Dimakis, and M.~Gastpar, ``Local interference can accelerate
  gossip algorithms,'' \emph{IEEE J. on Selected Topics in Signal Processing},
  vol.~5, no.~4, pp. 876--887, Aug 2011.

\bibitem{GoBo12}
M.~Goldenbaum, H.~Boche, and S.~Sta\'nczak, ``Nomographic gossiping for
  $f$-consensus,'' in \emph{Proc. of Intl' Symposium on Modeling and
  Optimization in Mobile, Ad Hoc and Wireless Networks (WiOpt)}, Paderborn,
  Germany, May 2012, pp. 130--137.

\bibitem{DiSa08}
A.~G. Dimakis, A.~D. Sarwate, and M.~J. Wainwright, ``Geographic gossip:
  Efficient averaging for sensor networks,'' \emph{IEEE Transactions on Signal
  Processing}, vol.~56, no.~3, pp. 1205--1216, Mar 2008.

\bibitem{MaHo74}
M.~Feinberg and F.~M. Horn, ``Dynamics of open chemical systems and the
  algebraic structure of the underlying reaction network,'' \emph{Chemical
  Engineering Science}, vol.~29, pp. 775--787, 1974.

\bibitem{DiSp07}
P.~Dittrich and P.~Speroni~di Fenizio, ``Chemical organization theory,''
  \emph{Bulletin of Mathematical Biology}, vol.~69, no.~4, pp. 1199--1231, May
  2007.

\bibitem{Fe72}
M.~Feinberg, ``Complex balancing in general kinetic systems,'' \emph{Archive
  for Rational Mechanics and Analysis}, vol.~49, no.~3, pp. 187--194, Dec 1972.

\bibitem{MeTh}
T.~Meyer, ``On chemical and self-healing networking protocols,'' Ph.D. Thesis,
  University of Basel, Switzerland, Feb 2011.

\bibitem{MeTs09a}
T.~Meyer and C.~Tschudin, ``Chemical networking protocols,'' in \emph{Proc. of
  the ACM Workshop on Hot Topics in Networks (HotNets)}, Oct 2009.

\bibitem{VaHo08}
A.~Varga and R.~Hornig, ``An overview of the {OMNeT++} simulation
  environment,'' in \emph{Proc. of SIMUTOOLS}, Marseille, France, 2008, pp.
  1--10.

\bibitem{DiZi01}
P.~Dittrich, J.~Ziegler, and W.~Banzhaf, ``Artificial chemistries - a review,''
  \emph{Artificial Life}, vol.~7, no.~3, pp. 225--275, Summer 2001.

\bibitem{Ab86}
H.~I. Abrash, ``Studies concerning affinity,'' \emph{J. of Chemical Education},
  vol.~63, no.~12, pp. 1044--1047, Dec 1986.

\bibitem{GiBr00}
M.~A. Gibson and J.~Bruck, ``Efficient exact stochastic simulation of chemical
  systems with many species and many channels,'' \emph{J. Phys. Chem. A}, vol.
  104, no.~9, pp. 1876--1889, Feb 2000.

\bibitem{Mc67}
D.~A. McQuarrie, ``Stochastic approach to chemical kinetics,'' \emph{J. of
  Applied Probability}, vol.~4, no.~3, pp. 413--478, Dec 1967.

\bibitem{JaHu07}
T.~Jahnke and W.~Huisinga, ``Solving the chemical master equation for
  monomolecular reaction systems analytically,'' \emph{J. of Mathematical
  Biology}, vol.~54, no.~1, pp. 1--26, Sep 2007.

\bibitem{Gi00}
D.~T. Gillespie, ``The chemical {L}angevin equation,'' \emph{J. Chem. Phys},
  vol. 113, no.~1, Jul 2000.

\bibitem{OlMu04}
R.~Olfati-Saber and R.~M. Murray, ``Consensus problems in networks of agents
  with switching topology and time-delays,'' \emph{IEEE Transactions on
  Automatic Control}, vol.~49, no.~9, pp. 1520--1533, Sep 2004.

\bibitem{ShRa13}
W.~Shaochuan and M.~G. Rabbat, ``Broadcast gossip algorithms for consensus on
  strongly connected digraphs,'' \emph{IEEE Transactions on Signal Processing},
  vol.~61, no.~16, pp. 3959--3971, Aug 2013.

\bibitem{ShAl09}
G.~Shinar, U.~Alon, , and M.~Feinberg, ``Sensitivity and robustness in chemical
  reaction networks,'' \emph{SIAM J. Appl. Math.}, vol.~69, no.~4, pp. 977 --
  998, 2009.

\bibitem{CaFa10}
R.~Carli, F.~Fagnani, P.~Frasca, and S.~Zampieri, ``Gossip consensus algorithms
  via quantized communication,'' \emph{Automatica}, vol.~46, no.~1, pp. 70--80,
  2010.

\bibitem{HoJa72}
F.~Horn and R.~Jackson, ``General mass action kinetics,'' \emph{Archive for
  Rational Mechanics and Analysis}, vol.~47, no.~2, pp. 81--116, 1972.

\bibitem{LeAn05}
P.~D. Leenheer and D.~Angeli, ``Monotonicity and convergence in chemical
  reaction networks,'' in \emph{Proc. of IEEE Conf. on Decision and Control,
  and European Control Conf. (CDC--ECC)}, Seville, Spain, Dec 2005, pp.
  2362--2367.

\end{thebibliography}

\end{document}